\documentclass[10pt, conference]{IEEEtran}

\usepackage{url}
\usepackage{multirow}
\usepackage{mdwlist}
\usepackage{mdwmath}
\usepackage{mdwtab}
\usepackage{cite}
\usepackage{array}
\usepackage{tabularx}
\usepackage[normalem]{ulem}
\usepackage{color}
\usepackage[ruled, vlined]{algorithm2e}
\usepackage{listings}
\usepackage{balance}
\usepackage[cmex10]{amsmath}
\usepackage{graphicx}
\DeclareGraphicsExtensions{.pdf,.jpeg,.png,.jpg}
\graphicspath{{./figures/}}
\usepackage{subfig}

\newcommand{\REM}[1]{}

\newtheorem{theorem}{Theorem}

\begin{document}

\title{JSweep: A Patch-centric Data-driven Approach for Parallel Sweeps on Large-scale Meshes}
\author{
\IEEEauthorblockN{Jie Yan$^{\star}$ \quad Zhang Yang$^{\dagger \ddagger}$ \quad Aiqing Zhang$^{\dagger \ddagger}$ \quad Zeyao Mo$^{\dagger \ddagger}$}
\IEEEauthorblockA{
$\dagger$ Software Center for High Performance Numerical Simulation, CAEP\\
$\ddagger$ Institute of Applied Physics and Computational Mathematics, Beijing, China\\
$\star$ Noah's Ark Lab, Huawei Technologies\\
Correspondence: \{yang\_zhang, zhang\_aiqing, mo\_zeyao\}@iapcm.ac.cn}
}
\maketitle

\begin{abstract}
In mesh-based numerical simulations, sweep is an important computation
pattern. During sweeping a mesh, computations on cells are strictly ordered by
data dependencies in given directions. Due to such a serial order,
parallelizing sweep is challenging, especially for unstructured and
deforming structured meshes. Meanwhile, recent high-fidelity
multi-physics simulations of particle transport, including nuclear reactor and
inertial confinement fusion, require {\em sweeps} on large scale meshes with
billions of cells and hundreds of directions.
In this paper, we present JSweep, a parallel data-driven computational
framework integrated in the JAxMIN infrastructure. The essential of JSweep is
a general patch-centric data-driven abstraction, coupled with a high
performance runtime system leveraging hybrid parallelism of MPI+threads and
achieving dynamic communication on contemporary multi-core clusters. Built on
JSweep, we implement a representative data-driven algorithm, Sn transport,
featuring optimizations of vertex clustering, multi-level priority strategy
and patch-angle parallelism. Experimental evaluation with two real-world
applications on structured and unstructured meshes respectively, demonstrates
that JSweep can scale to tens of thousands of processor cores with reasonable
parallel efficiency.
\end{abstract}

\IEEEpeerreviewmaketitle

\section{Introduction}
\label{sec:introduction}

In mesh-based numerical simulations, {\em sweep} is an important computation pattern widely used in solving deterministic Boltzmann Transport Equation (BTE)~\cite{boltzmann}, convection dominated or Navier-Stokes equation~\cite{dd-app-1-5}\cite{dd-app-2-2} and so on. During {\em sweep} on a mesh, cells are computed from upwind to downwind in the sweeping direction. One cell can compute only when all of its upwind neighboring cells are computed.

General {\em sweep} computation on large-scale meshes is challenging. For rectangular structured meshes where the data dependencies can regular, the well-known Koch-Baker-Alcouffe(KBA)~\cite{t3d}\cite{cm200} algorithm which uses a pipelining wavefront way to parallelize multiple sweeps has been very successful. However, for the more general deforming structured meshes and unstructured meshes in which data dependencies among cells are irregular, the KBA method doesn't fit and is typically impossible. Instead, a data-driven approach~\cite{plimpton-2000}\cite{mo}\cite{pautz} is often considered. This approach models the cells' data dependencies as a directed acyclic graph (DAG), regardless of the mesh types, then {\em sweep} on the mesh is equivalent to a topological traversal on the DAG. Unfortunately, although KBA-based sweep on regular structured meshes has scaled to $10^6$ CPU cores and billions of cells in 2015~\cite{parsec-sweep},  sweep on ordinary unstructured meshes still doesn't efficiently scale to $10^5$ cores. 

Meanwhile, mesh-based application programming frameworks\cite{jasmin}\cite{jaumin}\cite{samrai}\cite{pumi} have been increasingly critical to today's complex simulations that require to couple multiple multi-physics procedures. On one hand, multiple simulation procedures developed on a unified framework share the same specification of mesh and data layout, and thus are more consistent to work together. Given the fact that coupling different simulation procedures is difficult for both software development and numerical validation, this really makes sense. On the other hand, by providing users a programming abstraction and ensuring reasonable performance, the framework isolates applications from the evolution of underlying HPC systems, and thus achieves good portability. Recently, in areas related to particle transport in which sweep on mesh is the most time-consuming portion, framework-based coupling of multi-physics simulations have led to several successful cases, including the full core reactor simulation based on MOOSE\cite{multiphysics-reactor} and the ICF (Inertia Confinement Fusion) program LARED-I\cite{lared-all} based on JASMIN\cite{jasmin}. Nevertheless, so far these cases are still on structured meshes only.

In this paper, we focus on the patch-based mesh application framework, specifically JAxMIN~\cite{jasmin}\cite{jaumin} (detailed in section \ref{sub:jaxmin}), where the mesh is divided into patches. Patch is conceptually a subdomain of the mesh. The patch-based approach has advantages on adaptive mesh refinement, mesh management and parallel computation scheduling. Over 50 real-world applications have been implemented on JAxMIN. However, JAxMIN, like most counterparts, adopts BSP (Bulk Synchronous Parallel) \cite{bsp} style of parallel computing, in which all subdomains (patches) first compute using previous data of themselves and other subdomains, and then communicate to update their remote copies. Although being efficient and scalable enough for most numerical solvers, BSP is seriously inefficient for data-driven sweep computation where the parallelism is fine-grained. Furthermore, the fact that a patch often can't finish computation at one time and thus has to compute many times, as well as complex factors impacting performance, makes it hard to realize in JAxMIN's BSP-based abstraction.

\begin{figure}[h]
	\centering
	\includegraphics[scale=0.35]{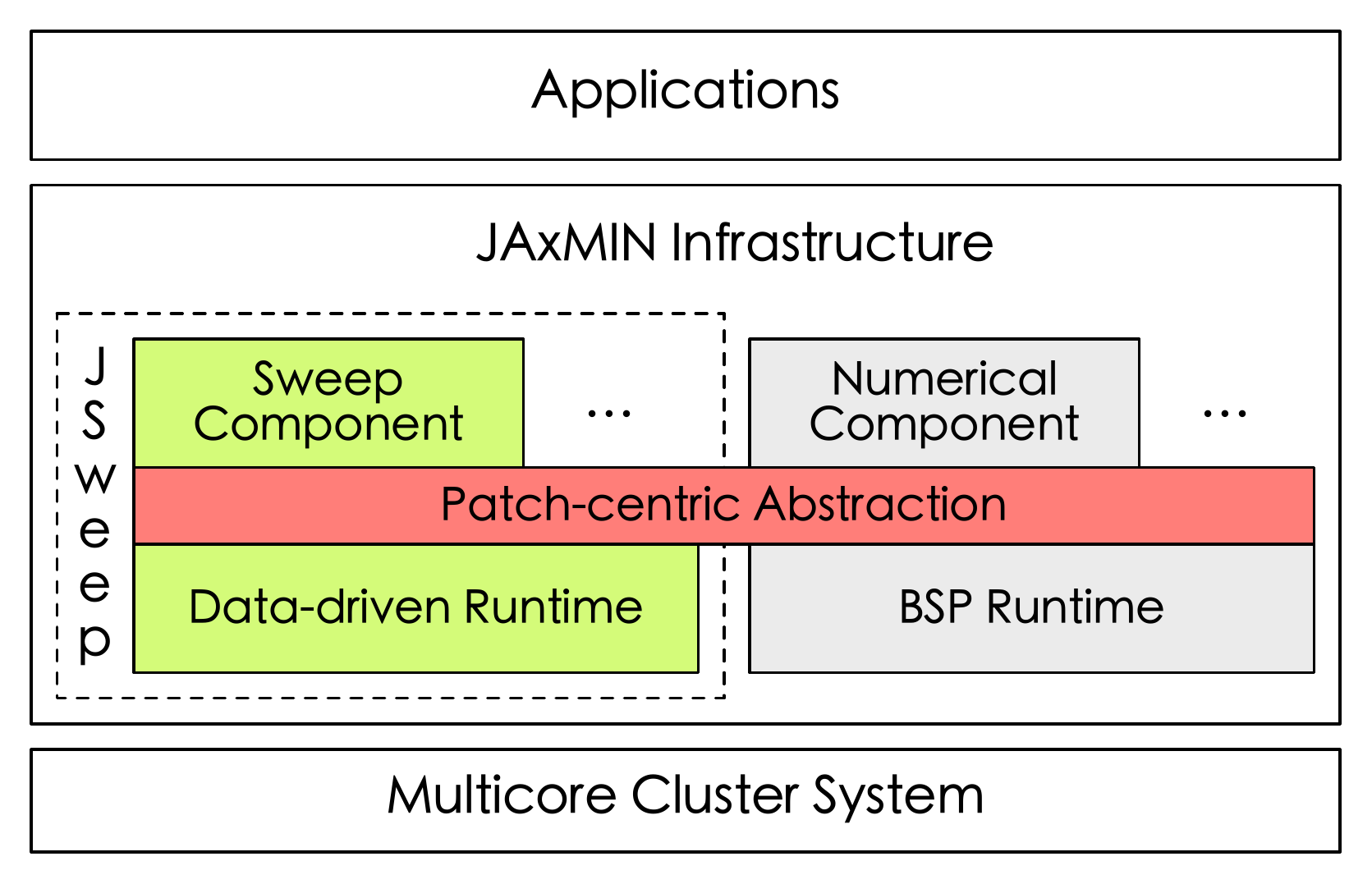}
\caption{JSweep Framework Overview}
\label{fig:jsweep}
\end{figure}

We propose JSweep, a patch-centric data-driven approach for parallel sweep computation on both structured and unstructured meshes, embedded in the JAxMIN infrastructure. Fig.\ref{fig:jsweep} illustrates the JSweep modules in the abstracted layers of JAxMIN. Specifically, our key contributions are as follows:
\begin{itemize}
\item The patch-centric data-driven abstraction, a unified model for general data-driven procedures on both structured and unstructured meshes. The core idea is extending the concept of patch to a logical processing element that supports reentrant computation and communicate with other patches (Sec.~\ref{sec:abstraction}). 
\item The patch-centric data-driven runtime module for contemporary multi-core cluster systems, featuring hybrid parallelism (MPI+threads) and dynamic data delivery (Sec.~\ref{sec:runtime}).
\item A sweep component based on the above patch-centric approach, enhanced by vertex clustering, multi-level priority strategy, patch-angle parallelism and coarsened graph. (Sec.~\ref{sec:sweep}).
\item Experimental evaluation with real applications of particle transport on both structured and unstructured meshes, demonstrating JSweep's reasonable performance and scalability on up to 76,800 processor cores (Sec.~\ref{sec:evaluation}).
\end{itemize}

Besides, we present background and motivation in Sec.~\ref{sec:background}, related work in Sec.~\ref{sec:relatedwork}, and finally conclusions in Sec.~\ref{sec:conclusion}.

\vspace{-0.2cm}

\section{Background and Motivation}
\label{sec:background}

\subsection{Preliminaries}
\label{sub:preliminaries}

\begin{figure}[h]
	\centering
	\includegraphics[scale=0.25]{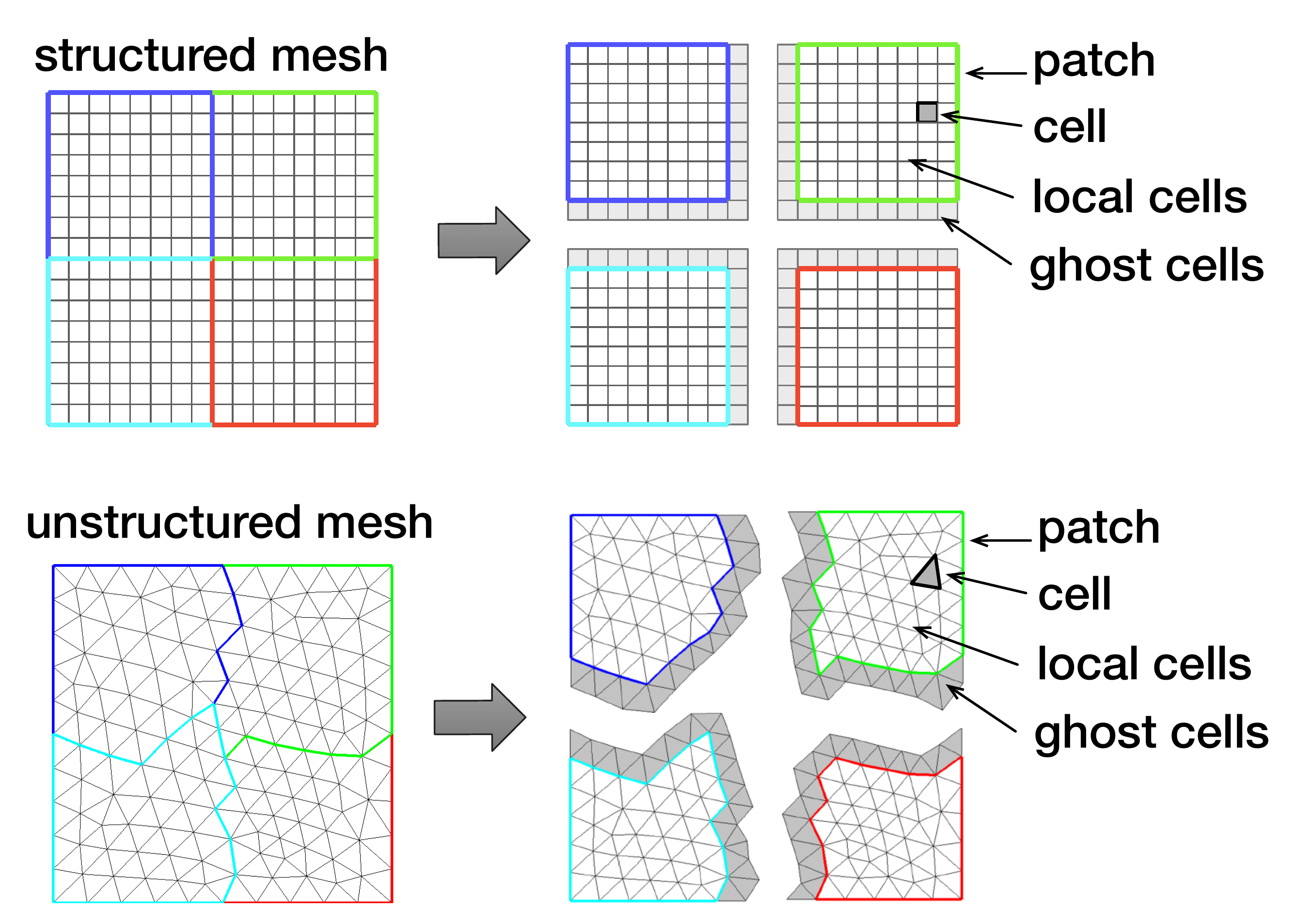}
\caption{Illustration of mesh terminologies.}
\label{fig:patch}
\end{figure}

Throughout this paper, we use a small set of terminologies as illustrated in Fig.~\ref{fig:patch}. Note that we describe them in an abstract view and don't explicitly differentiate the structured and unstructured meshes unless otherwise stated. 

\begin{itemize*}
\item{\tt mesh/grid:} the generic way of describing the discretized domain.
\item{\tt cell:} the smallest unit of a discretized domain.
\item{\tt patch:} a collection of contiguous cells.
\item{\tt local cells:} cells owned by a patch. They are updated by an operator applied to the patch.
\item{\tt ghost cells:} halo cells surrounding local cells. They are needed for computation but not updated by the local operator. They belong to other patches.
\end{itemize*}

\subsection{JAxMIN: Patch-based Mesh Application Infrastructure}
\label{sub:jaxmin}

JASMIN~\cite{jasmin} and JAUMIN~\cite{jaumin} (denoted as JA{\em x}MIN\footnote{Abbreviation of J Adaptive Structured/Unstructured Mesh INfrastructure.} for simplicity), are software infrastructures for programming applications of large-scale numerical simulations on structured and unstructured meshes. They share the same design principles, i.e., patch-based mesh management and component-based application programming interface. Although they are different in the way of describing mesh geometry and mesh elements, we omit these details by discussing in an abstract level in this paper.

JAxMIN adopts a patch-based strategy to manage the mesh and data. The computational domain, discretized as mesh, is decomposed into patches. {\em Patch} is a well-defined subdomain, that (1) each patch has complete information about its own cells as well as other mesh entities, (2) with ghost cells, each patch can explicitly get all adjacency information about its neighboring patches, and (3) it is abstract enough to hide differences of structured and unstructured meshes.

JAxMIN provides users a rich set of components as the programming interface. {\em Component} here is a generic implementation of any computational pattern. To implement a parallel program, users only need to instantiate a component by defining the application-specific computation kernel. So far, general patterns such as initialization, numerical computation, reduction, load rebalance, particle-in-cell communication, are provided. Besides, JAxMIN implements amounts of physics-specific integration components.

Traditionally, JAxMIN adopts the BSP model to organize computations in a component. The computations consist of a sequence of iterations, called super-steps. During a super-step, each patch executes logically in parallel, as follows: (1) does computation independently without data exchange with others, and then (2) does halo exchange communication with neighbors using newly computed data. Since many numerical algorithms fit well in BSP, the patch-based framework has been successful in many areas.

JAxMIN implements a high performance runtime system supporting hybrid MPI+threads parallelism and accelerators, with underlying optimization on memory management, data layout and buffering communication.

\subsection{Data-driven Parallel Sweeps}
\label{sub:sweep}

Without loss of generality, we consider the sweep computation in discrete ordinates ($S_n$) transport solvers. Sweep is the most computationally intensive portion of source iterative methods solving $S_n$ form of Boltzmann Transport Equation\cite{boltzmann}. As the name implies, {\em sweep} in any ordinate direction requires a computational ordering of cells from upwind to downwind. One cell can begin computing only if all of its upwind neighboring cells are computed.

Parallelizing sweep computation is challenging since it can't be efficiently implemented in a BSP manner. For regular structured meshes, the KBA approach, decomposing 3d meshes in a 2d columnar fashion and pipelining the computation for successive angles, is sufficient with BSP. However, for more general deforming structured meshes and unstructured meshes where data dependencies  among cells are irregular and thus the pipeline can't be easily determined, the KBA approach is almost impossible to implement.

\begin{figure}[h]
	\centering
	\subfloat{
		\includegraphics[scale=0.35]{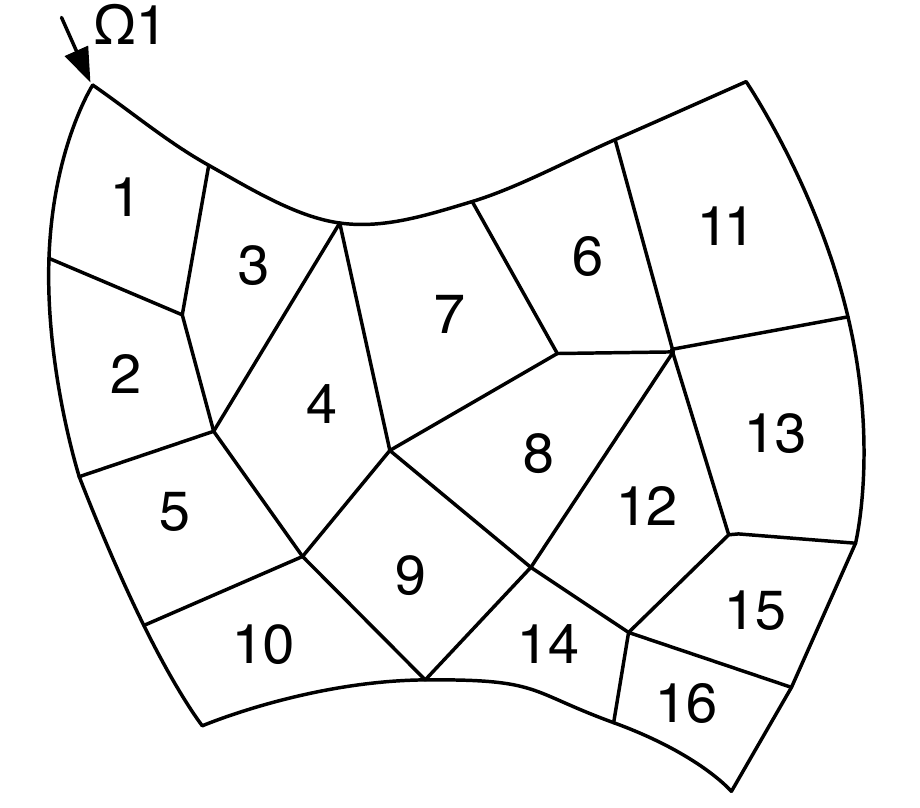}
		\label{subfig:grid}
	}
	\subfloat{
		\includegraphics[scale=0.35]{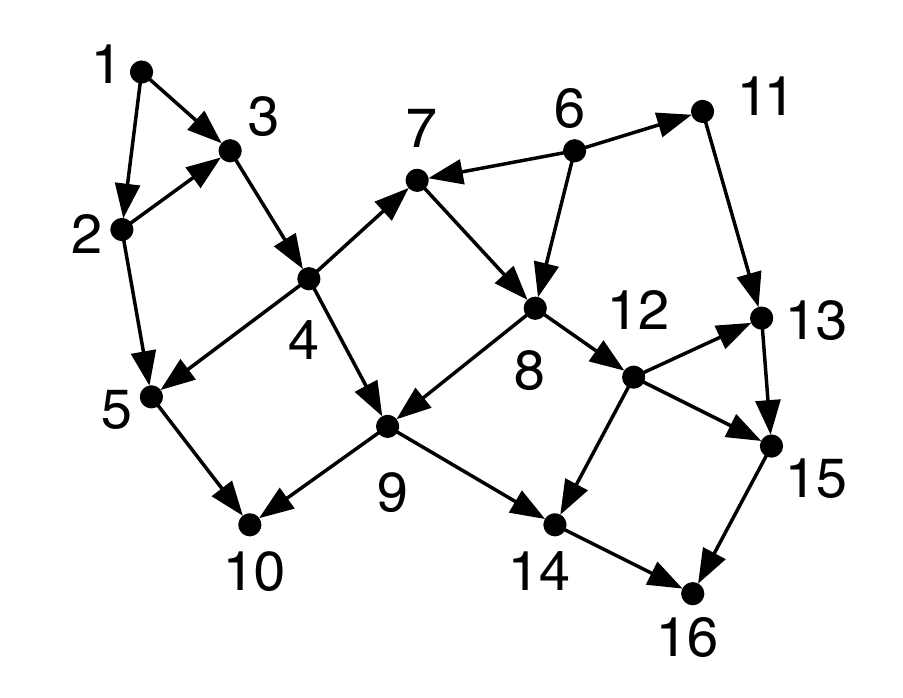}
		\label{subfig:dag}
	}
\caption{Illustration of sweeping an unstructured mesh from one direction and the induced data dependency graph\cite{plimpton-2000}}
\label{fig:mesh-and-dag}
\end{figure}

Alternatively, we focus on the data-driven parallelization which is a general approach for {\em sweeps} on both structured and unstructured meshes\cite{plimpton-2000}\cite{mo}. In this approach, any complex and irregular data dependencies can be explicitly modeled by a directed acyclic graph. As an example, Fig.~\ref{fig:mesh-and-dag} illustrates a 2d unstructured mesh and the associative directed graph in a given sweeping direction. Then, the {\em sweep} on a mesh is equivalent to a topological traversal on the directed graph, generalized with the user-defined numerical computations on the vertex.

In reality, multiple {\em sweep}s in different ordinate directions (angles) that are carried out simultaneously. We can model their induced data dependencies in a single graph and implicitly leverage parallelism of {\em sweeps} from all angles.

\subsection{Motivation}
\label{sub:motivation}
Now we consider the data-driven parallel sweep procedure in the context of the patch-based framework, especially JAxMIN. Unlike other numerical algorithms, patch-level data-driven computation can't be naturally supported in BSP, due to difficulties described below. These difficulties motivate us to develop a new patch-centric data-driven abstraction in the next section.

\subsubsection{Partial computation}
\begin{figure}[h]
	\centering
	\subfloat{
		\includegraphics[scale=0.35]{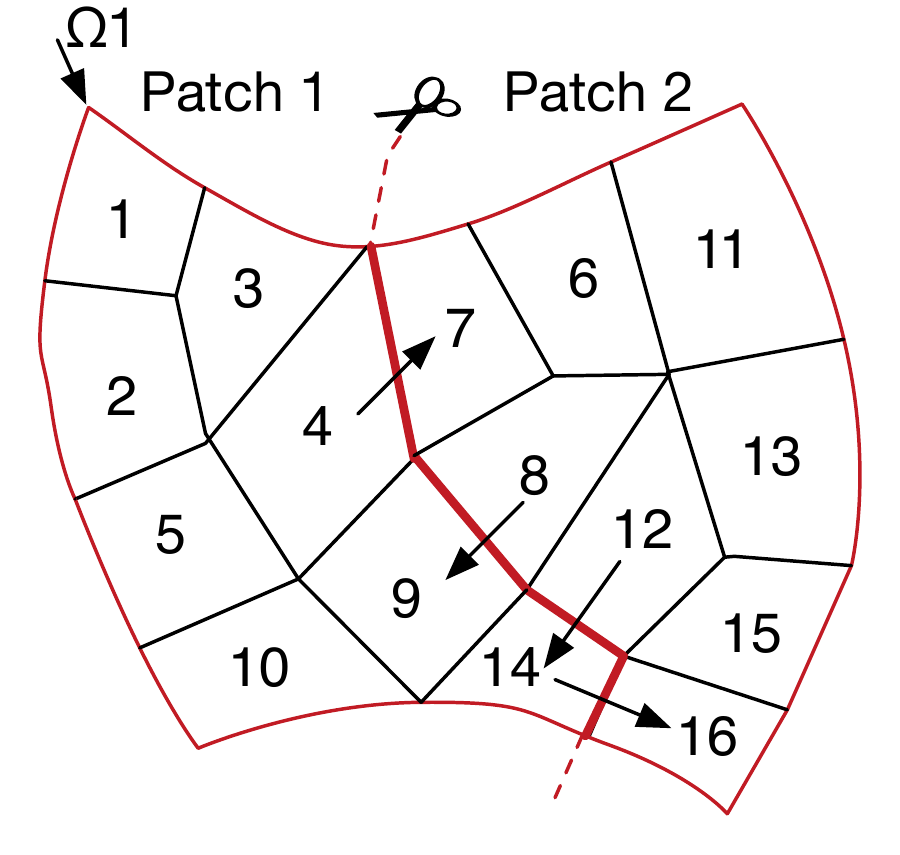}
	}
	\subfloat{
		\includegraphics[scale=0.35]{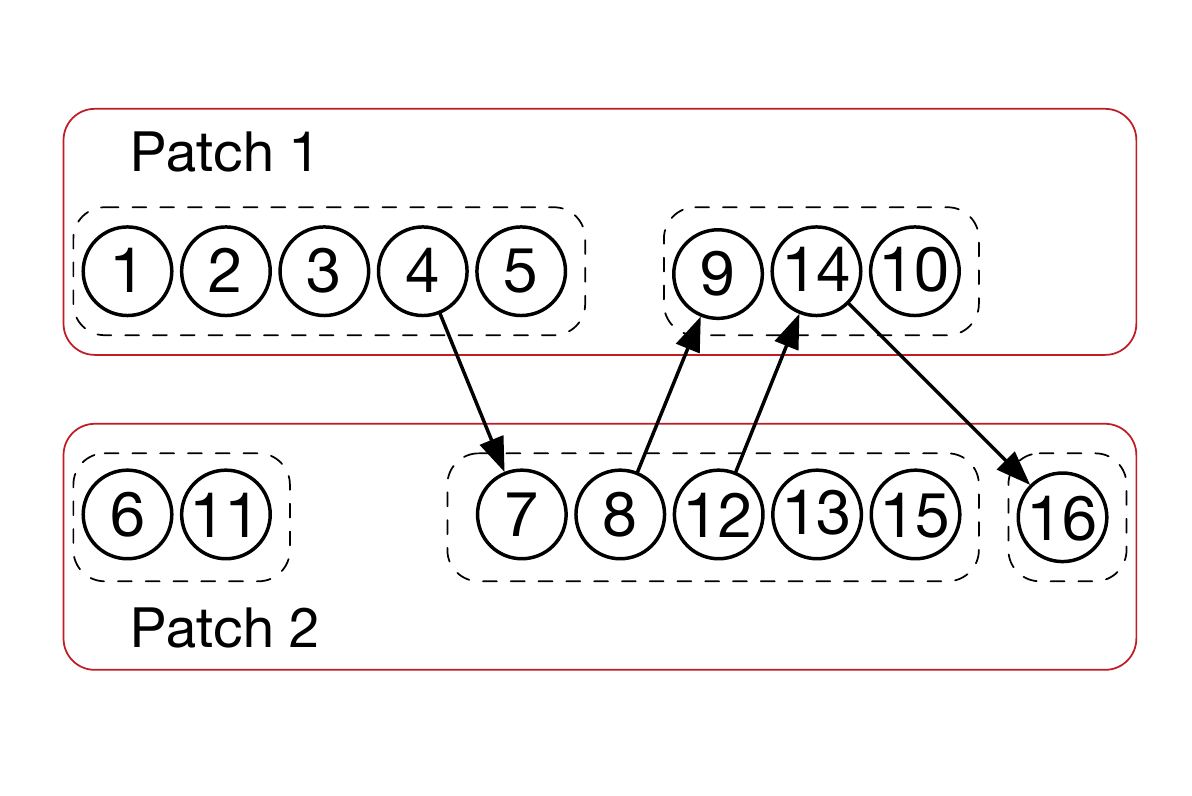}
	}
\caption{Illustration of sweeping on a 2d unstructured mesh decomposed into two patches (ghost cells are not shown).}
\label{fig:patches-sweep}
\end{figure}

In data-driven scenarios, partial computation of the patch is essential. As illustrated in Fig.~\ref{fig:patches-sweep} where one mesh is partitioned into two patches, interleaved data dependencies between the patches means that the patch can't be computed as a whole. In reality, the above zig-zag data dependency can be normal in unstructured meshes. Thus, to be reentrant, partial computation of a patch is necessary.

\subsubsection{Simultaneous sweeps on a patch}
\label{subsub:simultaneous-taks-on-patch}
\begin{figure}[h]
	\centering
  	\includegraphics[scale=0.4]{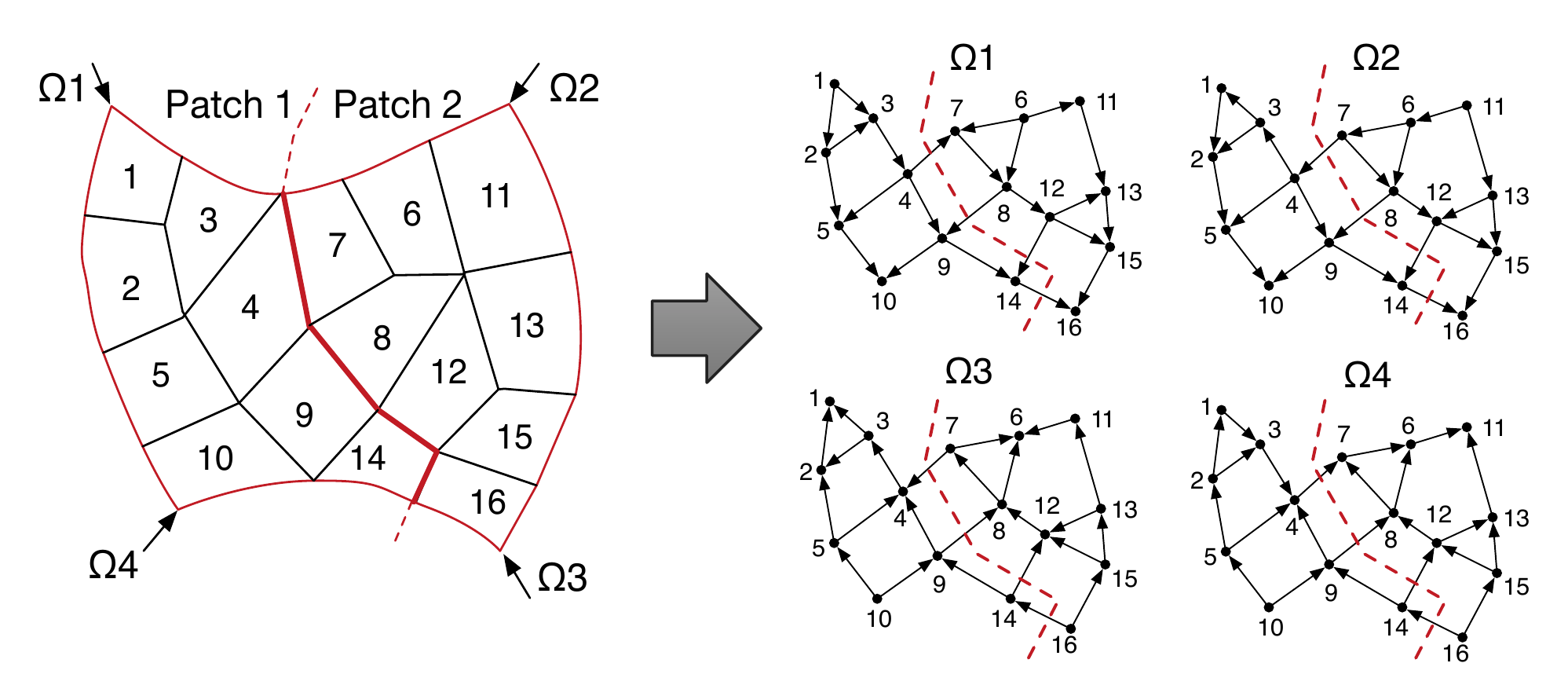}
\caption{Parallel sweeps from 4 independent directions(angles)}
\label{fig:sweeps-multi-angles}
\end{figure}

In real-world applications, sweeps from multiple directions are often performed in parallel. For example, in the $S_2$ sweeps example illustrated in Fig.~\ref{fig:sweeps-multi-angles}, one patch would be swept by multiple sweep procedures from 4 different directions. Generally, it is common that some sweeping directions are independent to each other. Thus, to enable such parallelism, simultaneous sweeps on a patch is necessary. In JAxMIN, however, patch is the basic unit of parallel computation, so we need to extend its abstraction.

\subsubsection{Priority strategies}
Priority strategies are becoming more important and complex. Previous work \cite{plimpton-2000}\cite{mo} have proven that ordering of computing cells (or vertices) is often critical to both parallelism and performance. In their settings, since one process (in MPI) or thread handles only one mesh subdomain, it is sufficient to compute priorities of cells (or vertices in the associated graph). However, in the patch-based framework, one process or thread typically is assigned with arbitrary number of patches, which means patch scheduling is always prioritized than cells within a patch. Thus, we need at least a two-level policy to prioritizing both patches and cells.

\section{Patch-centric Data-driven Abstraction}
\label{sec:abstraction}

In this section, we introduce the {\em patch-centric data-driven abstraction} for mesh-based parallel computations. Its foundation is the completeness and expressivity of the patch concept in JAxMIN described in section~\ref{sub:jaxmin}. In our abstraction, the concept of patch is further extended as a logical processing element being able to compute on itself and communicate with any other patches. Users should follow a {\em think-like-a-patch} philosophy to program, and focus on only actions of a single patch, i.e., defining the local computation and inter-patch communication. The abstraction doesn't expose any details of underlying patch execution details. It should be suitable for all patch-based mesh application frameworks including SAMRAI\cite{samrai}, (part-based) PUMI\cite{pumi} and especially JAxMIN.

\subsection{Data-driven patch-programs}
\label{sub:abstraction}
Data-driven logics on a patch is encoded as a {\em patch-program}. The patch-program is identified by a $(patch, task)$ pair, indicating $task$ is executed on $patch$. Any data communication between two patches is abstracted as a {\em stream}. The stream contains the user-defined data and description of source and dest patch programs. Fig.~\ref{fig:patch-program} presents the interface of patch-program and stream, in which the patch-program is factored into five primitive functions.

\begin{figure}[h]
{\sf
\lstset{language=c++, basicstyle=\small, tabsize=2, frame=single, numbers=none, numberstyle=\scriptsize, belowcaptionskip=4pt, xleftmargin=4pt, mathescape=true}
\lstset{emph={init, input, output, compute, vote_to_halt}, emphstyle=\color{red}, emph={[2]interface, struct}, emphstyle=[2]\color{green}}
\begin{lstlisting}
struct Stream {
	PatchID src_patch;//source patch
	TaskTag src_task; //task on source patch
	PatchID tgt_patch;//target patch
	TaskTag tgt_task; //task on target patch
	... //user-defined data
};
interface PatchProgram(PatchID $p$, TaskTag $t$) {
	void init();
	void input(Stream $s$);
	void compute();
	Stream output();
	bool vote_to_halt();
};
\end{lstlisting}
}
\caption{Patch-program interface}
\label{fig:patch-program}
\end{figure}

\vspace{-0.4cm}

\begin{figure}[h]
	\centering
	\includegraphics[scale=0.3]{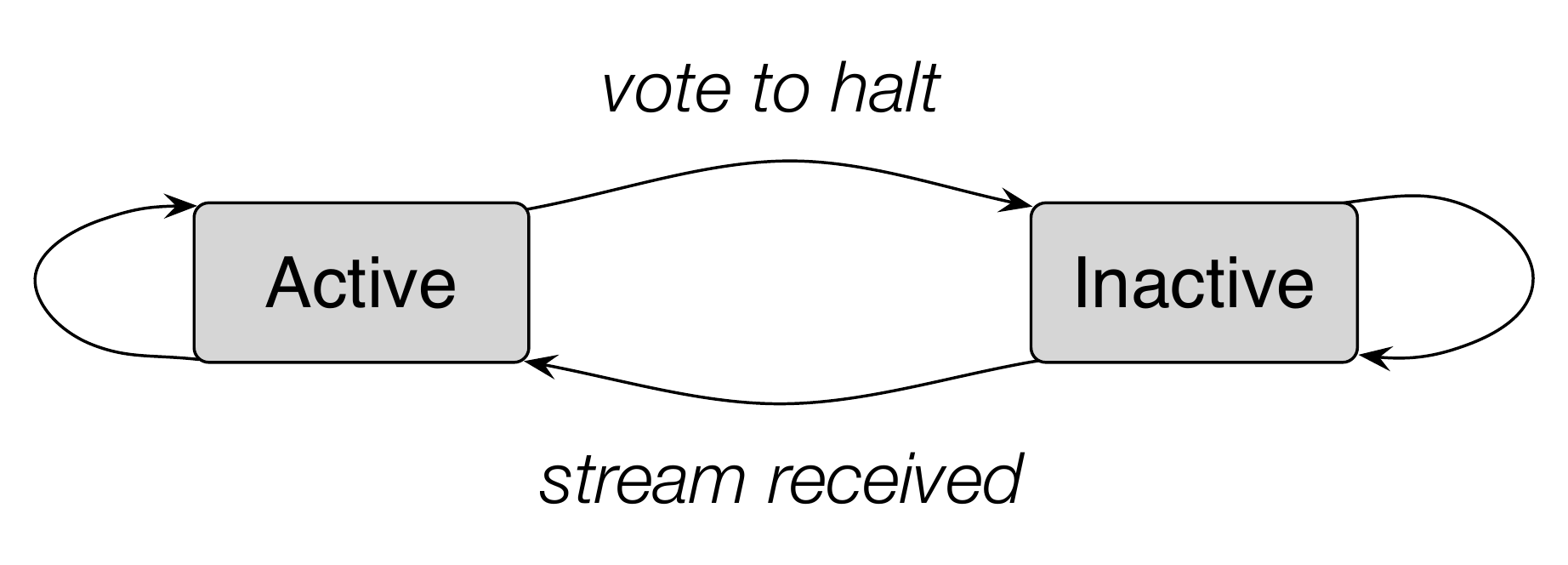}
\caption{Patch-program State Machine.}
\label{fig:psm}
\end{figure}

We define patch-program fully reentrant to support partial computation (detailed in the next subsection). At the beginning, each patch-programs is set {\em active}. And, in later execution, the state of a patch-program transits according to the finite state machine given in Fig.~\ref{fig:psm}. If its {\em vote\_to\_halt} function is evaluated true, the patch-program becomes {\em inactive}. Once receiving a {\em stream}, the patch-program becomes {\em active}. Conceptually, if there are no active patch-programs globally, the whole program terminates.

The patch-program, identified by ($patch = p, task tag = t$), is scheduled to run the semantics in Alg.~\ref{alg:patch-logics}, as follows.
\begin{itemize*}
\item If runs at the first time, the {\em init} function is used to initialize a local context.
\item Receives all streams sent to ($p, t$) by others, which is processed by the user-defined {\em input} function.
\item Calls {\em compute} function with user-defined numerical kernels.
\item Sends all {\em output} streams to and activates targets.
\item Calls {\em vote\_to\_halt} to evaluate whether there remains ready work to do. If not, deactivates itself. Otherwise, keeps active for being scheduled again.
\end{itemize*}

\begin{algorithm}[h]
\caption{Patch-Program Execution Semantics}
\label{alg:patch-logics}
\SetAlgoVlined
\SetKwInOut{Input}{input}
	\Input{Center patch $p$}
	\Input{Task tag $t$}
	\Begin{
\sf{
	\tcp{\color{blue} Init the state (execute once).}
	\nl \If{first time}
	{
		\nl {\color{red} init();}
	}
	\tcp{\color{blue} Recv data streams.}
	\nl \While{{\em Stream} s = {\sf receive}($p$, $t$) is not empty}
	{
		\nl {\color{red} input}{\tt ($s$)};
	}
	\tcp{\color{blue}  Compute.}
	\nl {\color{red} compute}{\tt ()};\\
	\tcp{\color{blue} Send data streams.}
	\nl \While{{\em Stream} s = {\color{red} \sf output}{\em ()}  is not empty}
	{
		\nl activate{\tt ($s.tgt\_patch$, $s.tgt\_task$)};\\
		\nl send{\tt ($s$)};
	}
	\tcp{\color{blue} Vote to halt.}
	{
	\nl \If{{\sf {\color{red} vote\_to\_halt}}{\em ()}}
	{
		\nl deactivate{\tt ($p$, $t$)};
	}
	}
}
}
\end{algorithm}

\subsubsection{Partial computation of patch-program}
\label{sub:partial-computation}
Partial computation is an essential property of the patch-program. On one hand, generally a patch-program couldn't finish at one time and thus requires many times of scheduling. As illustrated by the S$_n$ {\em sweeps} case (Sec~\ref{sub:motivation}), two patch-programs would depend on data of each other, leading to a dead lock if patch-programs are not reentrant. On the other hand, a patch-program may contain multiple parts of computations that depend on data of different patch-programs, so allowing a patch-program to execute multiple times can benefit from finer grained parallelism.

In our abstraction, {\em partial computation} of a patch-program is achieved by the following approaches. First, we allow storing of local context so the state are memorized, as illustrated by the implementation of {\em sweeps} in section~\ref{sub:sweep-basic}. Second, the logics of {\em finite state machine} in Fig.~\ref{fig:psm} maintains state transition of a patch-program, ensuring the correctness of termination after arbitrary times of partial execution.

\subsubsection{Simultaneous tasks on a patch}
\label{sub:multi-tasks}
Our abstraction supports multiple tasks on the same patch. Since any patch-program is identified by the pair ($patch, task$), multiple tasks on a patch naturally execute in parallel, even with possible inter-task communications. Whether and how to decompose work on a patch into patch-programs is the programers' decision. For the full S$_n$ sweeps discussed in Sec.\ref{subsub:simultaneous-taks-on-patch}, by defining sweep on any patch $p$ from the angle $a$ as a patch-program ($patch=p, task=a$), sweeps from all directions execute simultaneously.



\subsection{Scheduling patch-programs}
\label{sub:scheduling-patch-programs}
The data-driven engine initializes and continues to schedule active patch-programs to run until program termination.

For general patch-centric data-driven computations, the necessary and sufficient condition of program termination is that {globally \em all $(patch, task)$s become inactive}. To detect the termination condition in distributed situations, general negotiating protocols\cite{consensus} are needed. However, in numerical algorithms requiring the data-driven approach, the workload is known in advance. Thus, we can often detect the termination with little or even no distributed negotiation. For example, in {\em sweeps}, the program termination condition is all $(cell, angle)$s are computed, which is known by every patch before computation, and termination detection only need local information. In JSweep's real implementation, we actually allow the patch-program to commit its remained workload (i.e., number of $(cell,angle)s$ in {\em sweeps}) to a data structure shared by the master and worker threads of local runtime system (detailed in next section). The master thread, as representative of the process, participates distributed terminate negotiation only when there are no longer patch-programs with remained workload.


Priority policies are known critical for scheduling computations, yet is tightly coupled with the properties of the problem itself. In section~\ref{sub:opt-priority}, we shall discuss several strategies used in parallel $S_n$ {\em sweeps}.

\section{Patch-centric Data-driven Runtime System}
\label{sec:runtime}
\begin{figure}[h]
	\centering
	\includegraphics[scale=0.25]{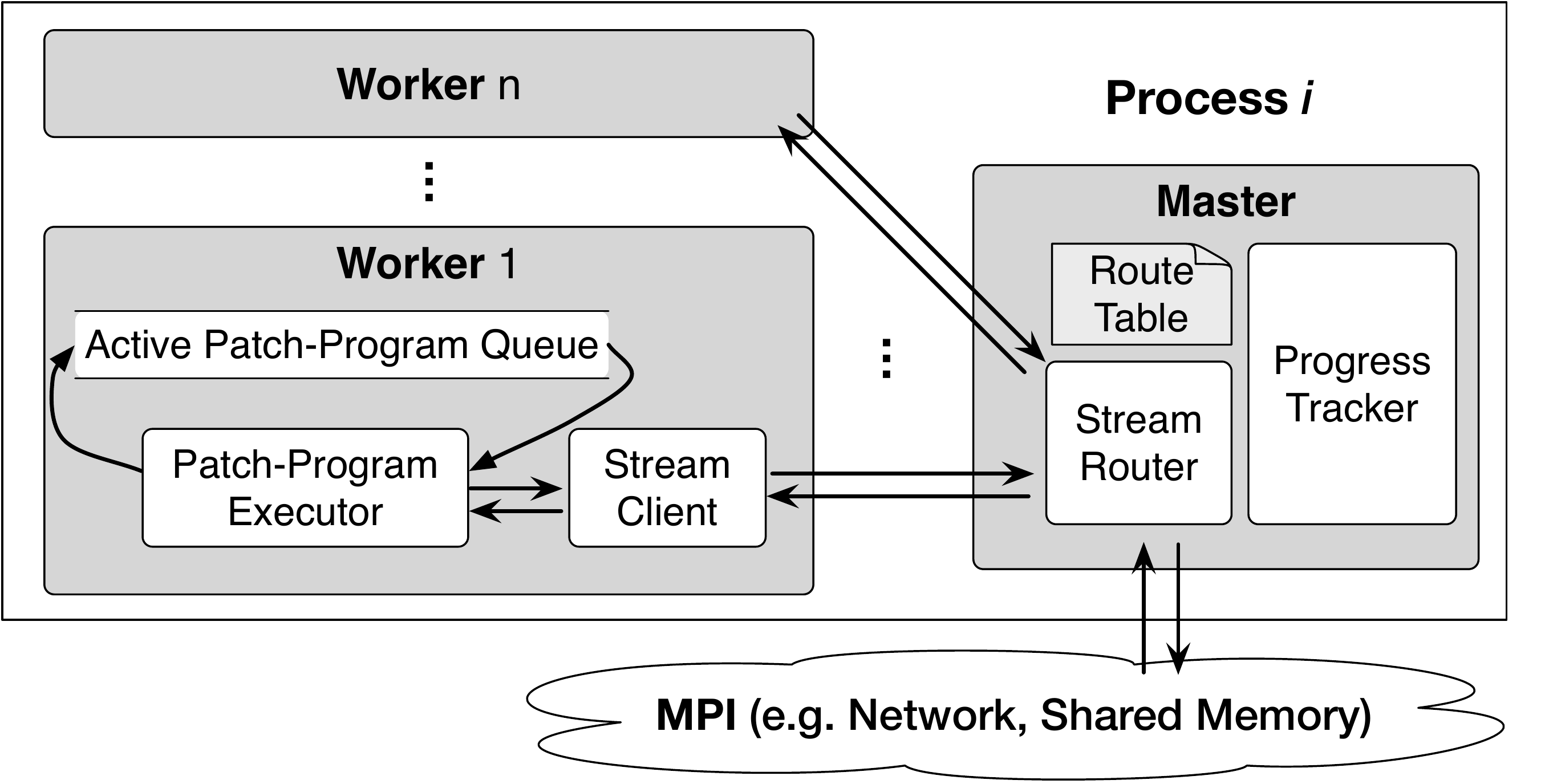}
\caption{Data-driven Runtime Overview}
\label{fig:rtm}
\end{figure}

In this section, we present the runtime that maps the patch-centric data-driven computation and communication onto underlying resources. Our target platform is the {\em multicore cluster} widely adopted in contemporary HPC systems. Fig.~\ref{fig:rtm} shows an overview of the runtime system. In particular, our design emphasizes fast stream delivery, load balance, fine-grained parallelism and low schedule overhead.
\vspace{-0.2cm}

\subsection{Hybrid parallelism}
\label{sub:rtm-parallelism}
The runtime inherits from JAxMIN a hybrid parallel approach of {\tt MPI + threads}, in which the program is organized with distributed-memory MPI processes and each process consists of multiple threads. On top of this design, JAxMIN have been highly optimized on both domain decomposition and data management. Besides, to reduce NUMA effect in multi-socket systems, by default our runtime launches one MPI process per processor and bind the process to the processor. Within each MPI process, the master-workers multi-threading mode is adopted. As shown in Fig.~\ref{fig:rtm}, the master thread is in charge of scheduling patch-programs, communicating streams and detecting global termination, while each worker thread executes patch-programs and communicates only with master thread.

\subsection{Dynamic stream communication}
\label{sub:rtm-communication}
In the patch-centric abstraction (section~\ref{sub:abstraction}), communication conceptually happens between a pair of $(patch, task)$s and is defined as a {\em stream}. For data-driven algorithms, the latency of {\em stream} transmission is critical for performance, since only received the dependent data can a patch-program execute. By definition, {\em the communication of stream is asynchronous and dynamic.}

The runtime system adopts the routable stream concept and reserves a specific core for master thread to support such timely communication. As defined in section~\ref{sub:abstraction}, the {\em stream} itself carries full information of source and target $(patch, task)$s. By identifying the target patch-program and looking up the route table that maps $(patch, task)$ to $(process, thread)$ , the runtime system can deliver any stream to its target place, either locally or in remote process.

The master thread schedules patch-programs by routing streams. At the beginning, all patch-programs are active and assigned to workers evenly. Later as the execution progresses, some patch-programs become inactive. If the master thread receives a stream whose target patch-program is inactive, it chooses and sets a lightest worker as the patch-program's owner, and then routes it the stream.

\subsection{Distributed progress tracking}
\label{sub:rtm-progress}
The master thread calls the {\em progress tracker} to detect global program termination. Once temporarily there is no longer work within the process, the progress tracker is activated. A consensus algorithms is implemented to detect distributed termination\cite{consensus}. Besides the general negotiating protocol, as discussed in section ~\ref{sub:scheduling-patch-programs}, for known data-driven algorithms special condition detection methods would be preferred for efficiency in practice. Currently, we support both.

\REM{
\subsection{Known Issues}
Currently, we adopt an one-time assignment policy on scheduling newly ready $(patch, task)$ and one $(patch, task)$ may require multiple times to finish its work, such that priority inversion among workers is possible. In this situation, one worker may hold many high priority tasks that leads to serious waiting of other workers. In the worst case, although in practice rarely happen, it may make the whole execution serial.}

\section{New Parallel Algorithm of Sn Sweeps}
\label{sec:sweep}
Now we describe a new parallel sweeps algorithm based on JSweep, the above patch-centric data-driven approach. Further, we explore four optimizations, including scheduling by $(patch,angle)$, vertex clustering, multi-level priority strategy, and coarsened graph, which are natural and efficient to implement thanks to the expressibility of the patch-centric abstraction.

\subsection{Patch-Program Implementation}
\label{sub:sweep-basic}
We assume that the mesh has been decomposed into patches with general spacial domain decomposition methods (for example, the METIS~\cite{metis} and Chaco~\cite{chaco} for unstructured meshes, Morton and Hilbert space filling curves for structured meshes). Each process is assigned with an arbitrary number of patches, shared by all its threads.

Formally, we define the directed graph induced by sweeping meshes as $G = (V, E)$, where each {\em vertex} is a $(cell, angle)$ pair, and each {\em edge} is directed data dependency between two vertices. An edge $(u, v)$ means vertex $v$ depends on vertex $u$'s data. For any patch $p$ and a sweeping direction $t$, we denote the induced subgraph as $G_{p,t} = (V_{p,t}, E_{p,t})$, where $V_{p,t}$ is the set of vertices (i.e., \{ $(cell, t)$ \}) and $E_{p,t}$ is the set of edges.

Listing~\ref{patch-sweep} presents the patch-centric implementation of parallel sweeps. As presented, the patch-program consists of two parts, i.e., local context and interface implementation.

\lstset{language=c++, basicstyle=\small, tabsize=2, frame=lines, numbers=left, numberstyle=\scriptsize, numbersep=1pt, numberblanklines=false, belowcaptionskip=4pt, xleftmargin=4pt, mathescape=true}
\lstset{emph={init, input, output, compute, vote_to_halt}, emphstyle=\color{red}, emph={[2]solve}, emphstyle=[2]\color{blue}}
\begin{lstlisting}[caption={Patch-Program of data-driven parallel sweeps},label=patch-sweep, firstnumber=1]
//$G_{p,t} = (V_{p,t}, E_{p,t})$ is the subgraph of patch $p$ with
//task tag (i.e., sweeping angle) $t$.
SweepPatchProgram(PatchID $p$, TaskTag $t$)
{
	// Part 1: Local Context
	int counts[$|V_{p, t}|$];
	PriorityQueue $Q$;
	Map<Pair<PatchID,TaskTag>,Stream> $outstreams$;

	// Part 2: Interface Implementation
	void init() {
		$Q$.clear();
		for(each Vertex $v$ in $V_{p, t}$) {
			counts[$v$] = #. $v'$s upwind vertices;
			if(counts[$v$]==0) $Q$.enqueue($v$);
		}
	}
	void input(Stream $s$) {
		while((edge$(u, v)$,data($u$)) = $s$.read()) {
			counts[$v$] = counts[$v$]-1;
			if(count[$v$]==0) $Q$.enqueue($v$);
		}
	}
	void compute() {
		Vector<Vertex> $vertices$;
		// N is the vertex clustering grain
		while(!$Q$.empty() and $vertices$.size()<N) {
			Vertex $v$ = $Q$.dequeue();
			vertices.push_back($v$);
			for(each $v's$ downwind Vertex $w$) {
				if($w$ is in $V_{p, t}$) {
					counts[$w$] = counts[$w$]-1;
					if(counts[$w$]==0) $Q$.enqueue($w$);
				} else {
					Stream& $s$ = $outstreams$(patch($w$),$t$);
					s.write(edge($v$,$w$), data($v$));
				}
			}
		}
		solve($vertices$); //user-defined computation
	}
	Stream output() {
		Stream $s$ = fetch($outstreams$);
		return $s$;
	}
	bool vote_to_halt() {
		return Q.empty();
	}
}
\end{lstlisting}

The local context contains all necessary states required by a reentrant sweep on the patch, including: (line 6) an array of counters that count the number of unfinished neighbors for each local vertex, (line 7) a priority queue storing ready vertices, and (line 8) streams later sent to other patch-programs.

The interface functions implement DAG-based data-driven sweeps on the patch $p$ in the direction $t$. The {\em init} function initializes each vertex's count variable to the number of its upwind neighbors, and collect source vertices into the ready queue $Q$. The {\em input} function receives data of vertices from remote patches, updates counts of related local vertices; once a local vertex's count decreases to zero, put it to the ready queue. The {\em compute} function collect a sequence of ready vertices and computes on them with user-defined numerical computation, updates their downwind neighboring vertices. The {\em output} function generates streams sent to remote patch-programs. The {\em vote\_to\_halt} function evaluates whether the patch-program should deactivate.

\subsection{Optimization: Patch-Angle Parallelism}
\label{sub:opt1}
JSweep naturally supports simultaneous sweeps on a patch, from different angles (i.e., sweeping directions). As shown in Listing~\ref{patch-sweep}, we achieve this by setting task tag of the patch-program to the id of the sweeping direction. Consider the example in Fig.~\ref{fig:sweeps-multi-angles} again, in which full $S_2$ transport sweeps are carried on a 2d unstructured mesh of 2 patches. In this example, sweeps from different angular directions are independent, and thus patch-angle parallelism can be fully enabled. This is especially useful for small meshes with large number of angles.

\REM{
In particular, lets's focus on the colored patch (say $p$) and sweeping directions. The sweeps in $\Omega 1$ and $\Omega 2$ directions are from the same corner and induce same data dependencies between cells. In full sweeps, there are no dependencies between sweeping directions, thus patch-program($p$, $\Omega 1$) and patch-program($p$, $\Omega 2$) can be execute in parallel, which could reduce the idle time of computing resources (processor cores).

For sweeps Schedule by patch is not always efficient, since for a patch there is no {\em schedule locality} among different sweeping angles. Actually, the interval between starting time of two angles maybe very long, especially when the grid scale is large.
}

\subsection{Optimization: Vertex Clustering}
\label{sub:opt2}
We adopt vertex clustering in the patch-program. As shown in Listing~\ref{patch-sweep}, the {\em compute} function collects and computes on multiple ready vertices, rather than a single vertex. For example, in Fig.~\ref{fig:patches-sweep} vertices in the same dashed rectangular are clustered together.

Benefits of this optimization is two-fold. On one hand, vertex clustering can dramatically reduce the scheduling overhead by reducing execution times of a patch-program. For example, in Fig.~\ref{fig:patches-sweep} the patch 1 and patch 2 need only two and three executions respectively, compared to eight times of no clustering. On the other hand, vertex clustering aggregates multiple streams to the same target into a single stream and thus reduces communication overhead. For example, in Fig.~\ref{fig:patches-sweep}, the inter-vertex communications of $8 \rightarrow 9$ and $12 \rightarrow14$ can be combined into one message.

\begin{figure}[h]
	\centering
	\subfloat[vertex clustering]{
  	\includegraphics[scale = 0.2]{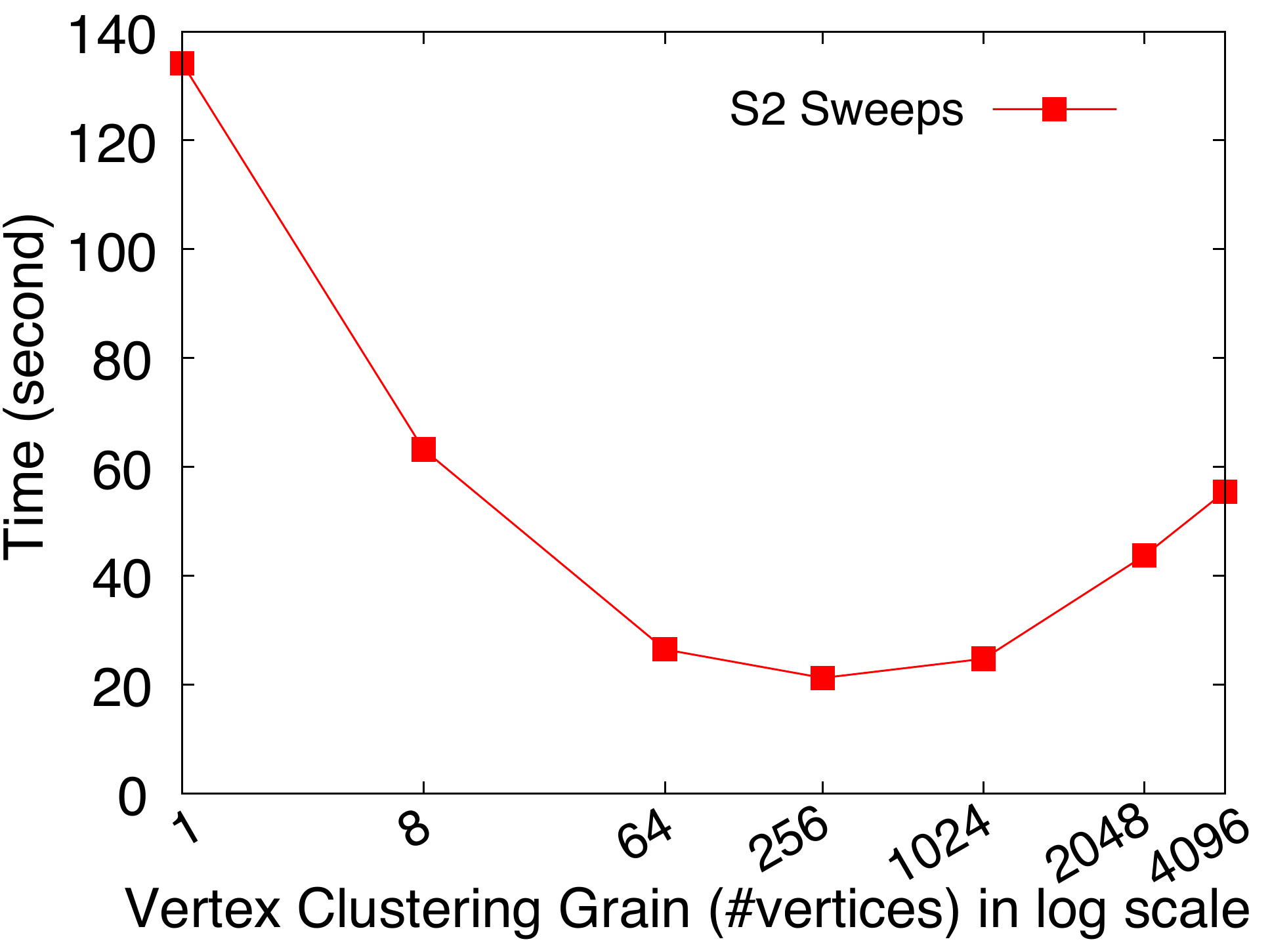}
  	\label{fig:snsweep-cluster-size}
	}
	\subfloat[priority strategy]{
  	\includegraphics[scale = 0.2]{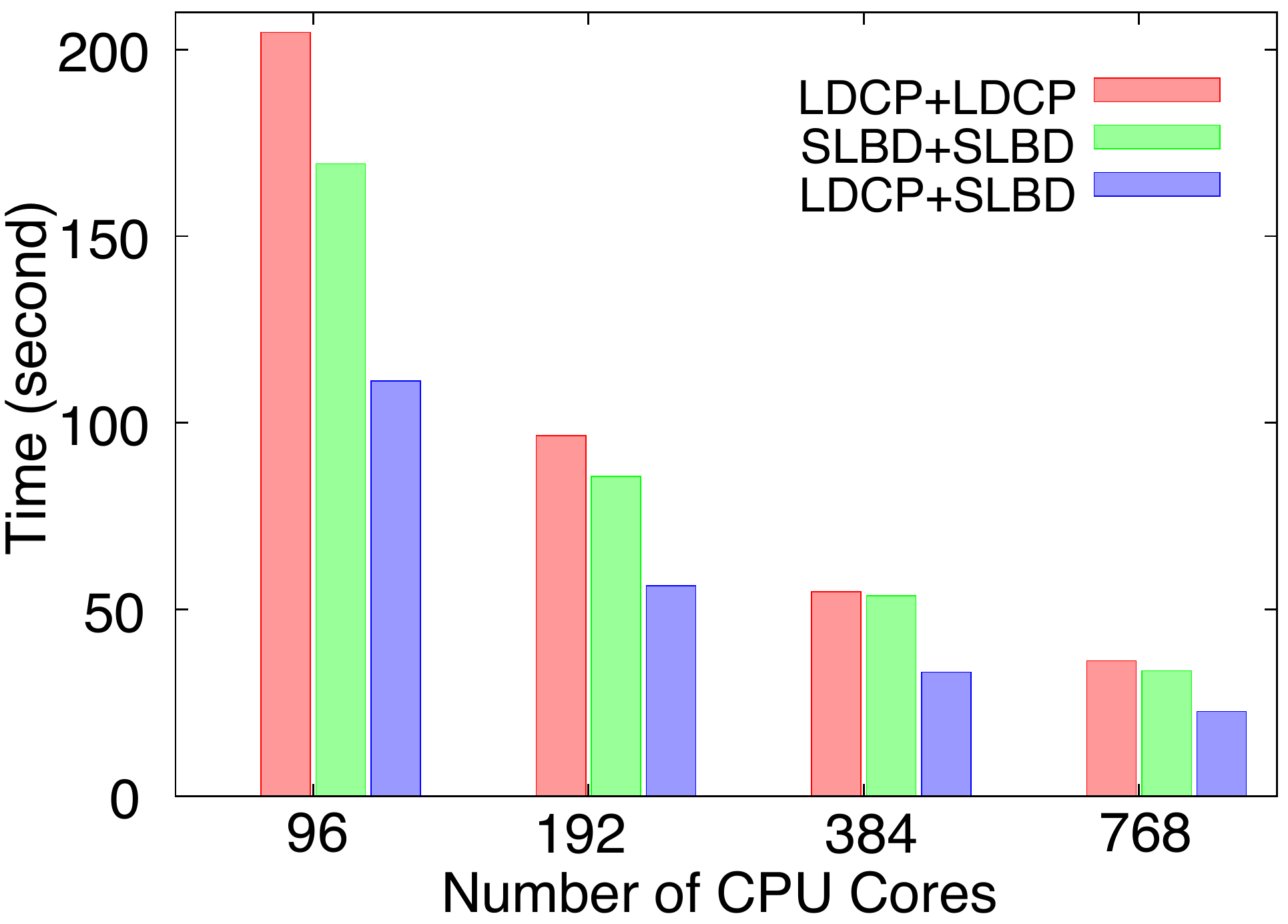}
  	\label{fig:snsweep-priority}
	}
\caption{Performance effect of optimization parameters}
\label{fig:performance-effect-s}
\end{figure}

Nevertheless, we need to choose the clustering grain carefully. While reducing overhead of schedule and communication, vertex clustering also has potentially negative effect on parallelism since it may defer the communication thus delay scheduling of other patch-programs. Excessive clustering can lead to long communication delay and thus longer execution time. To illustrate this, consider SnSweep-S, an example in JAxMIN package, which implements a $Sn$ solver for neutron transport equations on 3d structured meshes. The experimental results (mesh cells: $160 \times 160 \times 180$, patch size: $20 \times 20 \times 20$, S$_2$ ordinates, 8*12 CPU cores) are shown in Fig.~\ref{fig:snsweep-cluster-size}.

\vspace{-0.2cm}
\vspace{-0.2cm}

\subsection{Optimization: Priority Strategy}
\label{sub:opt-priority}
We adopt a two-level hierarchical priority strategy, i.e., $(patch, angle)$ priority and vertex priority.

The $(patch, angle)$ priority is used for JSweep runtime to schedule patch-programs. For patch $p$ and angle $a$, its priority is calculated by the following formula:
\\$~~~~~~prior(p, a) = prior(a)*C + prior(p)$
, where $C$ is a constant factor. In $S_n$ sweep, to avoid waiting of downwind patches, we want patch-programs with the same angle are continuously scheduled to execute such that the data streams are delivered to the nearby patches as quickly as possible. Thus we set the importance of $prior(a)$ always higher than $prior(p)$ in the formula, by multiplying a large factor $C$ over $prior(a)$. Meanwhile, with respect to $prior(p)$, however we can't reduce a single objective. On one hand, we hope the upwind patches are computed as earlier as possible such that more parallelisms are available. On the other hand, we also want the patches neignbouring other unfinished patches are computed earlier, but these preferred patches hare typically on the downwind of a sweeping direction. Based on the first objective, we develop two priority strategies: LDCP (Longest Distance on Critical Path) for structured meshes and BFS (Breadth First Search) for unstructured meshes. Based on the second objective, we develop the priority strategy SLBD (Shortest Local Boundary Distance, a DFS variant that prefers vertices most close to patch boundary) for both structured and unstructured meshes.

The vertex priority is used within a patch-program to order local ready vertices in PriorityQueue $Q$ in Listing.\ref{patch-sweep}. As $prior(p)$, vertex priority also has to trade off more parallelism and earlier communication. The strategies proposed for $prior(p)$, i.e., BFS, LDCP (for structured meshes only) and SLBD, are also suitable for vertex priority. In practice, however we observed that SLBD performs constantly best for unstructured and especially unstructured meshes, as illustrated by SnSweep experiments in Fig.\ref{fig:snsweep-priority}.


\subsection{Extra Optimization: Coarsened Graph}
\label{sub:opt-cg}

\begin{figure}[h]
	\centering
	\includegraphics[scale=0.35]{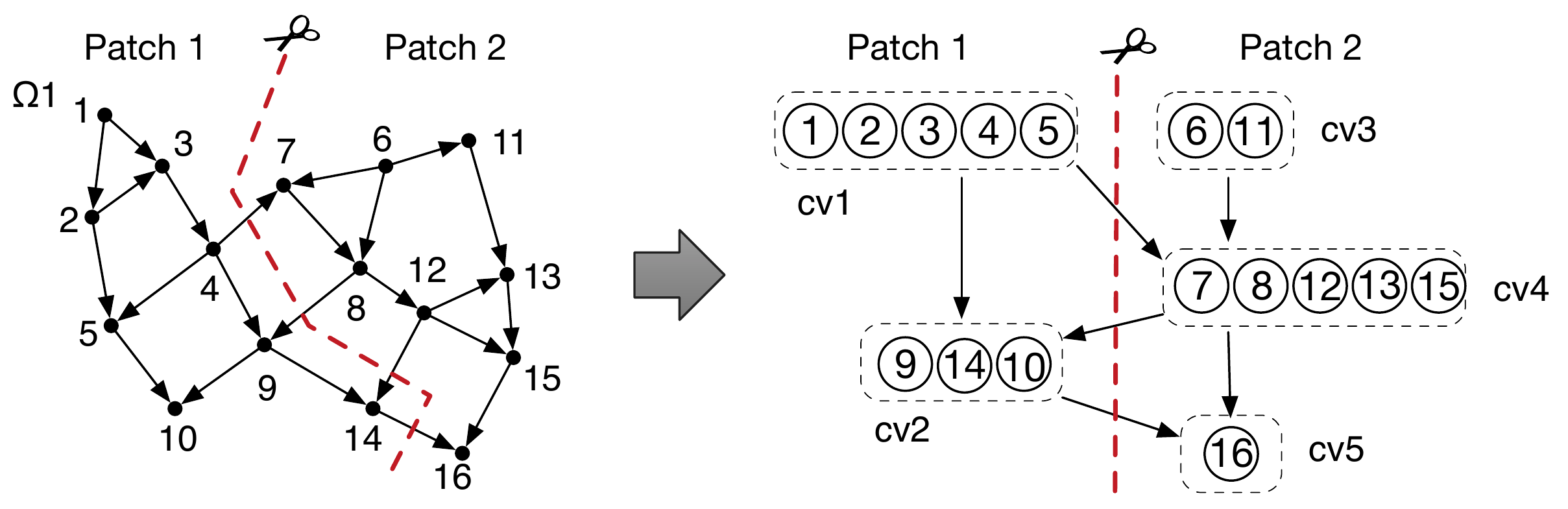}
\caption{Graph Coarsenning}
\label{fig:cg}
\end{figure}

Coarsened graph, not presented in Listing.\ref{patch-sweep} for length limit, can be treated as an extension to vertex clustering. In reality, the mesh structure and its data dependencies are always constant in most or even all sweeping iterations. Thus we can cache the vertex clustering results to build a reusable coarsened graph. For example, in Fig.~\ref{fig:cg}, the directed graph (left) is transformed into a much smaller coarsened graph (right) according to the previous clustering results in Fig.~\ref{fig:patches-sweep}.

Formally, we define a coarsened graph as the property graph: $\mathcal{C}G = (\mathcal{C}V, \mathcal{C}E, P(\mathcal{C}V), P(\mathcal{C}E))$ where $\mathcal{C}V$ is set of coarsened vertices derived and $\mathcal{C}E$ the set of coarsened edges. $\mathcal{C}G$ is the topology of vertex clusters and the directed communication relationships of vertex clusters. Property of a coarsened vertex $cv \in \mathcal{C}V$, $P(cv)$, is the series of corresponding clustered DAG vertices, while property of a coarsened vertex $ce \in \mathcal{C}E$, $P(ce)$, is the combined edges of source vertices and target vertices in DAG. For example, in Fig.~\ref{fig:cg}, $P(cv2) = (9,14,10), P(cv4)=(7,8,12,13,15), P(cv4 \to cv2)=(\{8,12\}, \{9,14\})$. Since $\mathcal{C}G$ is the task graph generated in the scheduling process, we have the computability theorem:

\begin{theorem}
If a directed graph $G$ is acyclic, its derived coarsened graph $\mathcal{C}G$ is also acyclic.
\end{theorem}

Besides, $\mathcal{C}G$ is distributed at the beginning of its construction. We implement it in same technologies presented in agent-graph~\cite{graphine}. With coarsened graph, sweep is carried on DAG in the first iteration and on $\mathcal{C}G$ in all subsequent iterations until the mesh changes. In our practice with JSNT-S\cite{jsnt-s}, the cost of building $\mathcal{C}G$ is less the one DAG-based sweep iteration itself while the speedup of sweeps on $\mathcal{C}G$ over DAG can be $7-10$ folds.

\REM{
\subsection{Application Interface}
\label{sub:app-interface}
Parallel {\em sweeps} has been implemented as a component in JAxMIN. Like other JAxMIN components, its application interface includes the following two parts.
\begin{itemize*}
\item {\tt patch strategy} class. Users need to define it following a class template which includes three primary member functions: (1) calculate data dependencies of local cells, (2) register patch data variables,  and (3) sweep on a given set of $(cell, angle)$ pairs. By defining this class, users can implement desired numeric computation, such as flux sweeping in neutron and radiation transport.
\item {\tt sweep integrator component} class. It requires the above user-defined {\tt patch strategy} as a parameter. In the application, users build a sweep component, and calls its member function to solve the sweeping computation in the mesh.
\end{itemize*}

The SnSweep examples in JAxMIN release packages illustrate how to define and use sweep and other components to implement source iterative method solving neutron transport equations on 3d structured and unstructured meshes.
}

\section{Evaluation}
\label{sec:evaluation}
{\bf Platform~} All experiments were carried on Tianhe-II, the world's fastest supercomputer in 2015\cite{top500}. We use at most 3200 nodes. Each node has two Intel Xeon E5-2692v2 12-core processors, equipped with 64GB memory and Tianhe-Express-II network of 40GB/s bandwidth. The operating system is Kylin Linux. All applications are compiled with Intel C Compilers (icc13) and customized MPICH2.

\begin{figure}[h]
	\centering
	\subfloat[Cube(structured)]{
  	\includegraphics[scale=0.4]{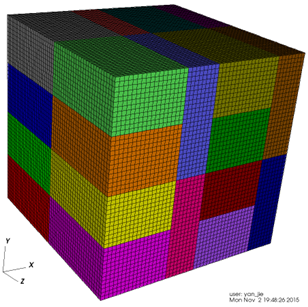}
  	\label{subfig:cube}
	}
	\hspace{0.3cm}
	\subfloat[Reactor(unstructured)]{
  	\includegraphics[scale = 0.32]{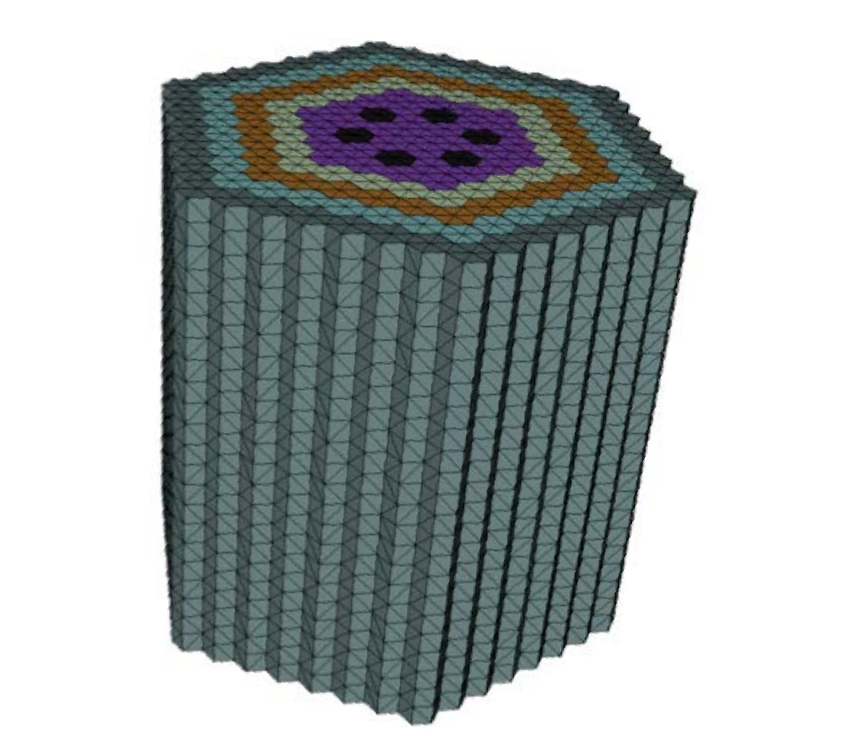}
  	\label{subfig:reactor}
	}
	\hspace{0.2cm}
	\subfloat[Ball(unstructured)]{
  	\includegraphics[scale=0.4]{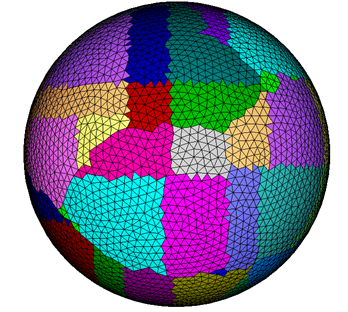}
  	\label{subfig:ball}
	}
\caption{Shapes of tested meshes}
\label{fig:input-jsnt}
\end{figure}

{\bf Applications~} We use two real JAxMIN-based Sn applications, JSNT-S~\cite{jsnt-s} and JSNT-U~\cite{3dsn}, to investigate the efficiency of JSweep on structured and unstructured meshes respectively. The used meshes are visualized in Fig.~\ref{fig:input-jsnt}.

\REM{
\begin{table}[htpb]
\caption{Dataset description}
\label{tab:meshes}
\centering
{\small
\begin{tabular}{|l|r|l|l|l|}
\hline
Name & \#cells & \#angles & \#groups & Other\\
\hline
Kobayashi400 & 400*400*400 & S16 & NA & NA\\
Kobayashi800 & 800*800*800 & S16 & NA & NA\\
\hline
Reactor & 64,479 & S4 & 4 &NA\\
Ball-I & 482,248 & S4 & 4 &NA\\
Ball-II & 173,197,768 & S4 & 4 &NA\\
\hline
\end{tabular}
}
\end{table}
}

\subsection{Evaluation on structured meshes}
\label{sub:evaluation-structured-mesh}
JSNT-S~\cite{jsnt-s} is a JASMIN-based S$_n$ package for structured meshes, which implements most functionalities of TORT~\cite{tort}. We use the well-known Kobayashi benchmark to evaluate JSweep. In particular, we focus on the strong scalability. In all the following experiments, JSweep is configured as follows: patch size = $20 \times 20 \times 20$, vertex clustering grain = 1000,  and the priority strategy is SLBD+SLBD.

\begin{figure}[h]
	\centering
	\subfloat[Middle scale (Kobayashi-400)]{
  	\includegraphics[scale=0.2]{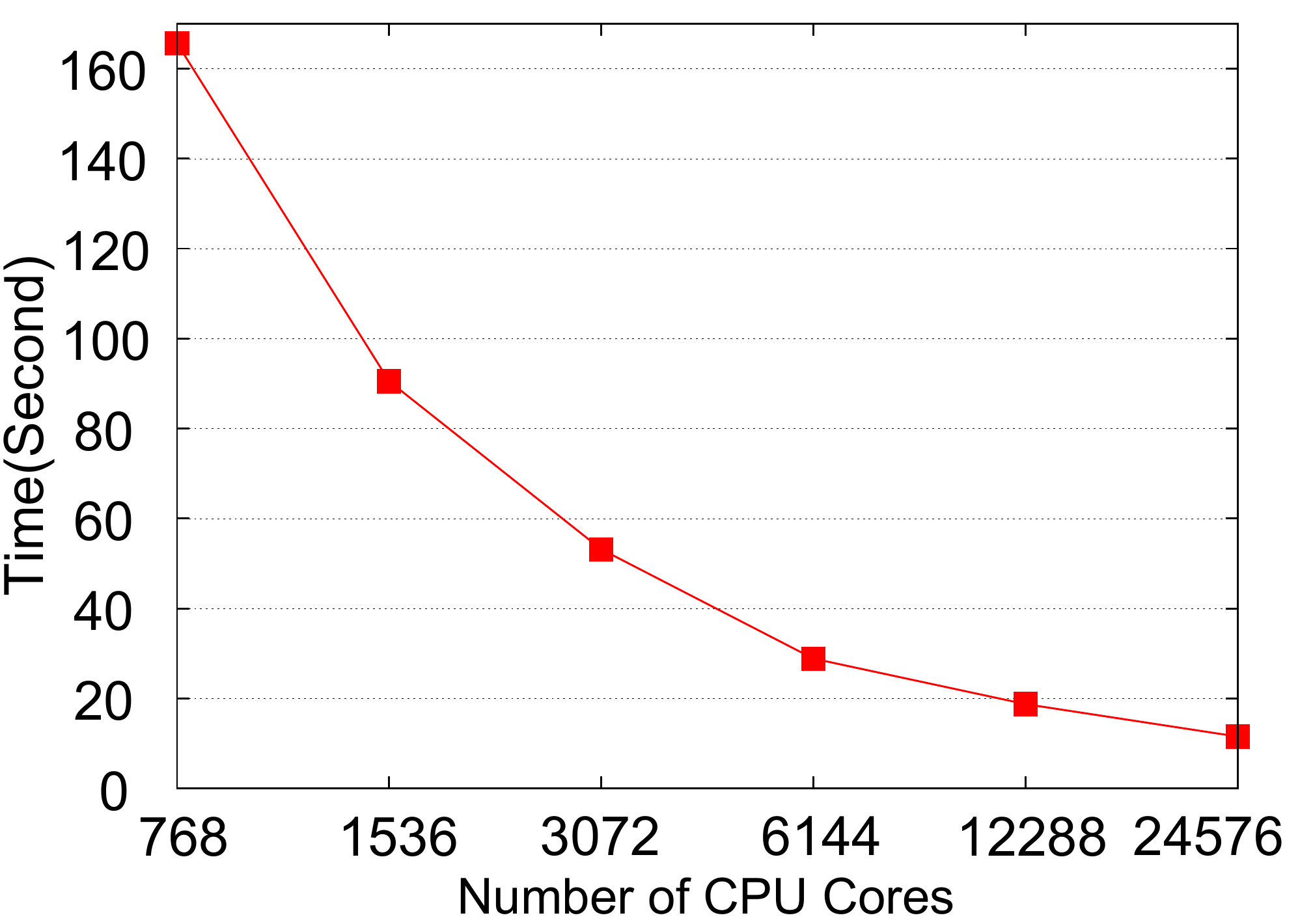}
  	\label{subfig:koba400}
	}
	\subfloat[Large Scale (Kobayashi-800)]{
  	\includegraphics[scale = 0.2]{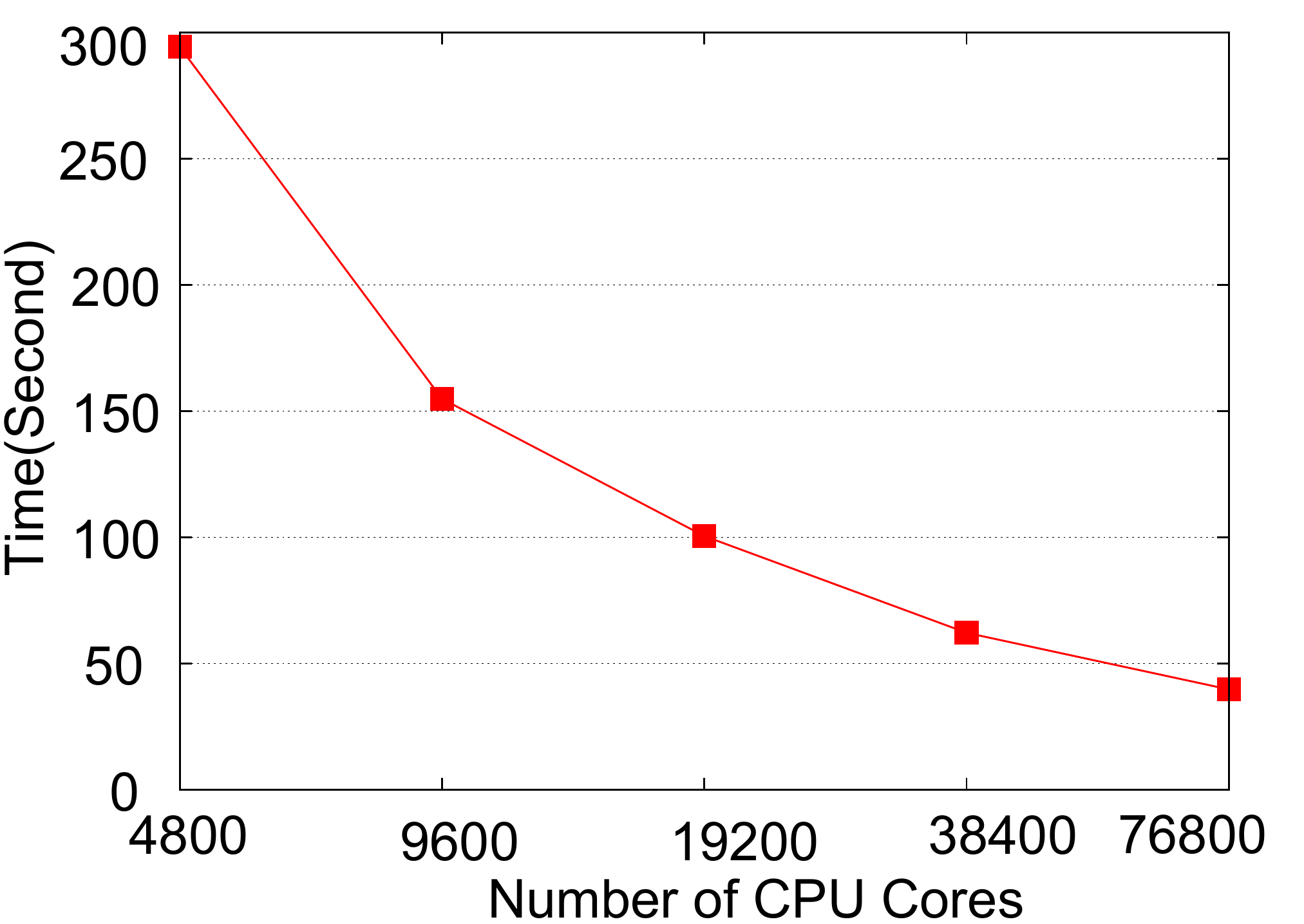}
  	\label {subfig:koba800}
	}
\caption{Runtime of JSNT-S for Kobayashi Benchmark}
\label{fig:jsnt-s-koba}
\end{figure}

We first evaluate JSweep with the original Kobayashi benchmark (Kobayashi-400). It solves the single energy group S$_n$ transport equations with scattering, on a cubic mesh (Fig.\ref{subfig:cube}) of $400 \times 400 \times 400$ cells with 320 angular directions. As presented in Fig.\ref{subfig:koba400}, with increasing number of cores, JSweep shows reasonable scalability constantly, with a speedup 14.3 (or parallel efficiency 44.7\%) on 24,576 cores compared to 768 cores. 

\vspace{-0.2cm}
\vspace{-0.2cm}

We then evaluate JSweep on more CPU cores with a larger problem by modifying the mesh of Kobayashi input to $800 \times 800 \times 800$ proportionally, namely Kobayashi-800. As shown in Fig.~\ref{subfig:koba800}, JSweep scales to 76,800 cores with a reasonable speedup 7.4 (or parallel efficiency 46.3\%), normalized to performance on 4,800 cores.

\REM{
Note that with respect to weak scalability whose results is not shown here, however, parallel efficiency of JSweep is not as good as KBA-based algorithms as reported in other literatures~\cite{denovo}\cite{ardra}. We conclude the reasons to two aspects. First, rather than the KBA's 2d columnar decomposition, we use a general domain decomposition methods which minimizes the volume-surface rate rather than optimize for pipelining sweeps. Second, the data-driven algorithm introduces overhead of DAG operations that is less scalable. For the second problem, we have proposed a DAG coarsening algorithm in another work~\cite{sweep-cg}, which dramatically reduces the DAG scale and improves the sweeps performance. }

\subsection{Evaluation on unstructured meshes}
\label{sub:evaluation-unstructured-mesh}
JSNT-U\cite{3dsn} is a JAUMIN-based S$_n$ package for unstructured meshes, primarily used for numerical simulations in high energy physics. We evaluate JSweep on two shapes of unstructured meshes, reactor core (Fig.\ref{subfig:reactor}) and ball (Fig.\ref{subfig:ball}). Unless otherwise stated, default configurations of experiments are as follows: priority strategy SLBD+SLBD, patch size = 500 cells, vertex clustering grain = 64, \#angles = 24 ($S_4$) and \#energy groups = 4.

\subsubsection{Hyper-parameters' effect to performance}
\begin{figure}[h]
	\centering
	\subfloat[Path size and cluster grain]{
  	\includegraphics[scale = 0.2]{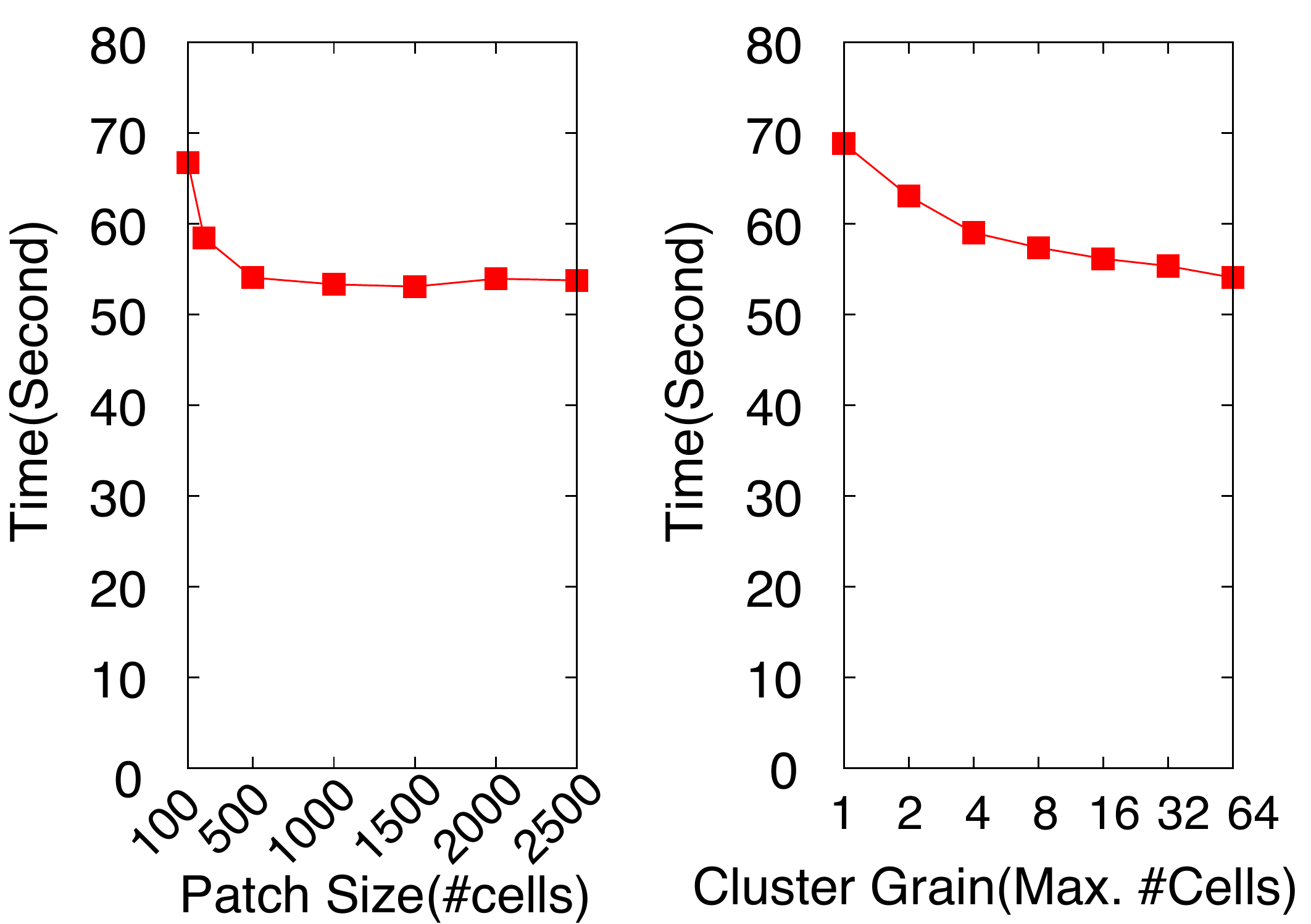}
  	\label{fig:jsnt-u-parameters}
	}
	\subfloat[Priority strategies]{
  	\includegraphics[scale = 0.2]{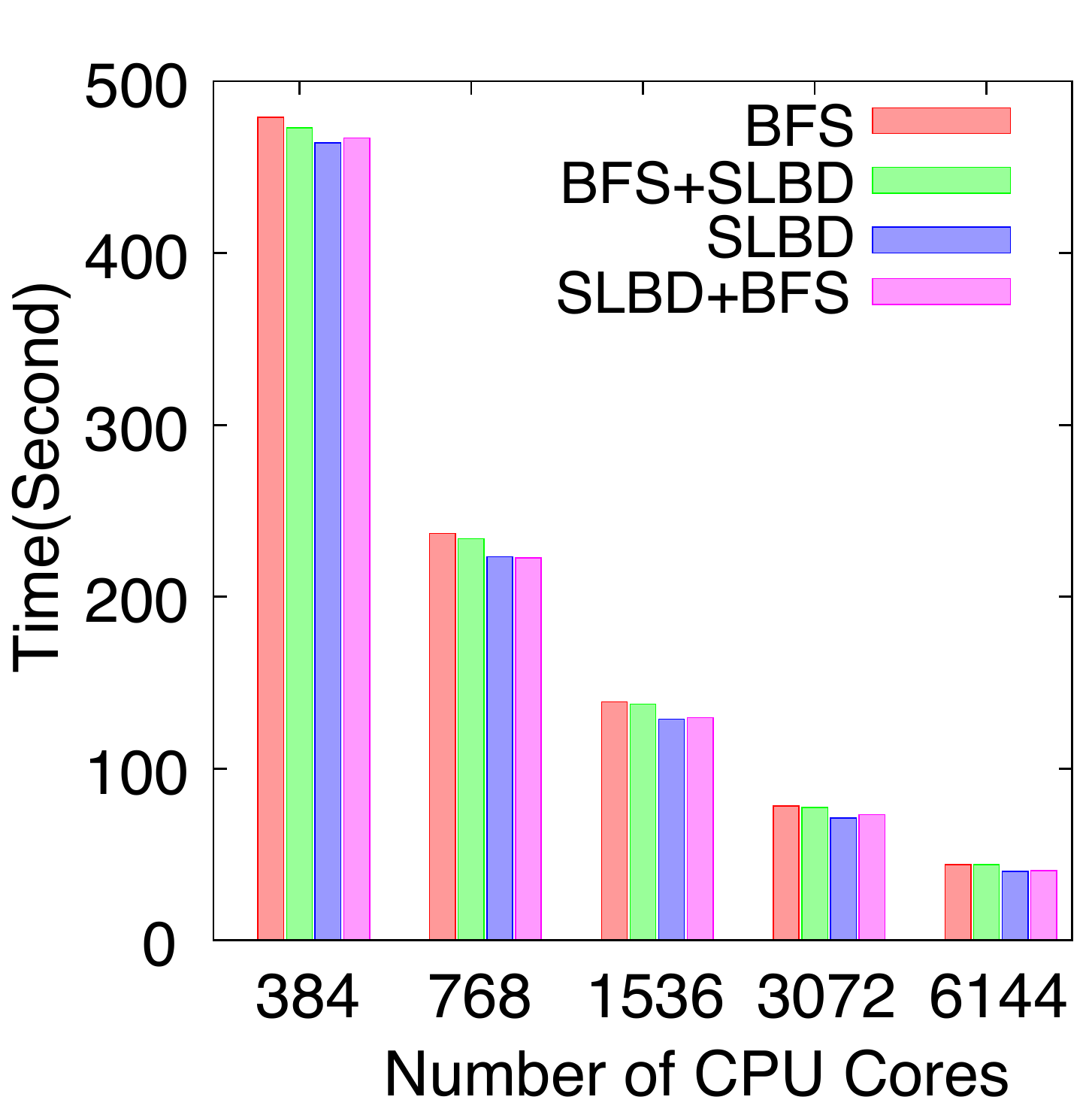}
  	\label{fig:jsnt-u-strong-priors}
	}
\caption{Hyper parameter's effect in JSNT-U (mesh: reactor)}
\label{fig:jsnt-u-parameter-effect}
\end{figure}

We change and investigate three hyper parameters respectively in order, i.e., patch size, vertex clustering grain and priority strategy, while keeping others default. As shown in Fig.\ref{fig:jsnt-u-parameters} (left), with increasing patch sizes (i.e., \#cells of a patch), the runtime first decreases quickly since the larger patch size reduces total communication between patches, and then slightly increases since the larger patch size also leads to longer waiting time of downwind patches whose execution is driven by data from this patch. Fig.\ref{fig:jsnt-u-parameters} (right) shows the effect of maximum vertex clustering grains. With increasing vertex clustering grain, the runtime decreases quickly and then keeps steady. Unlike on structured meshes (Fig.\ref{fig:snsweep-cluster-size}), however, the runtime no longer increases with a large clustering grain. By profiling, we found that the actual number of available vertices is between 16 and 64 at most time, which means the real clustering grain is limited by parallelism. With respect to priority strategies, as shown in Fig.\ref{fig:jsnt-u-strong-priors}, their effect to performance is not so significant as that on structured meshes (Fig.~\ref{fig:snsweep-priority}).
\vspace{-0.2cm}
\vspace{-0.2cm}

\subsubsection{Strong Scalability}
\begin{figure}[h]
	\centering
	\subfloat[Small scale (482,248 cells)]{
  	\includegraphics[scale = 0.2]{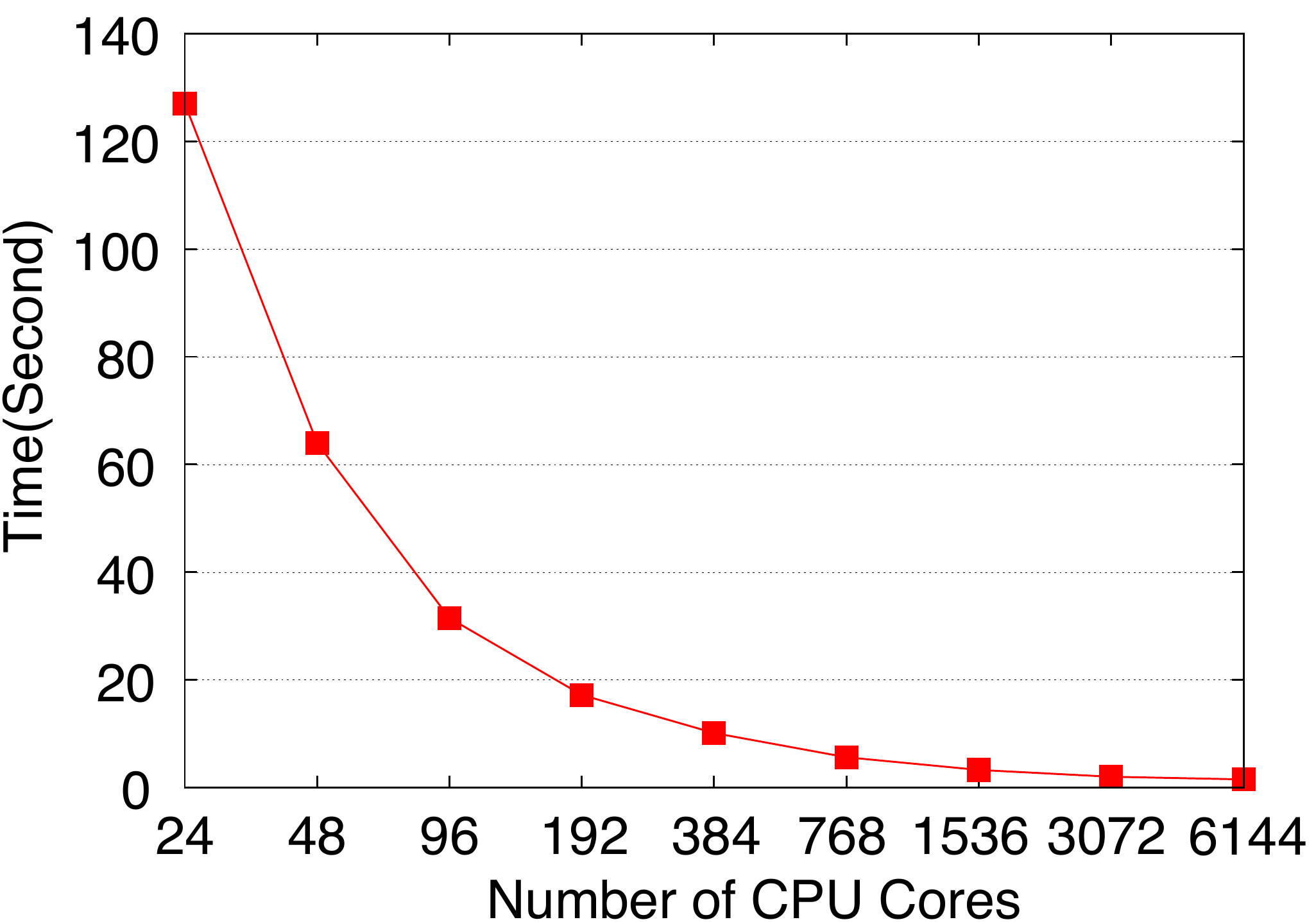}
  	\label{fig:jsnt-u-strong-r0}
	}
	\subfloat[Large scale (173,197,768 cells)]{
  	\includegraphics[scale = 0.2]{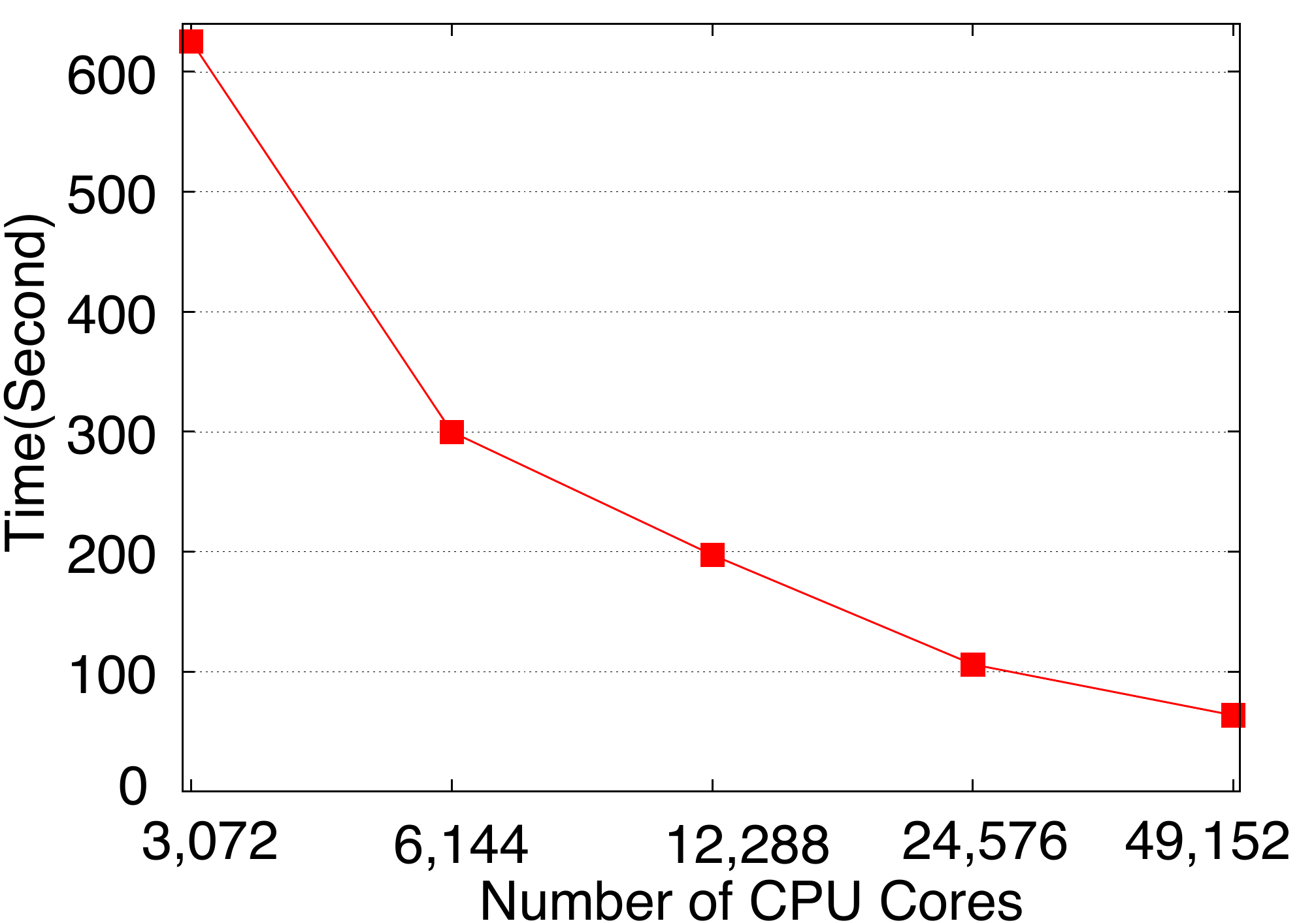}
  	\label{fig:jsnt-u-strong-r3}
	}
\caption{Strong scalability of JSNT-U on ball meshes}
\label{fig:jsnt-u-strong-scalability}
\end{figure}

As shown in Fig.~\ref{fig:jsnt-u-strong-scalability}, JSweep performs good strong scalability on both small and large meshes. For the small scale problem (ball of 482,248 cells), JSweep shows a speedup of 11.5 (parallel efficiency 72\%) at 384 cores and goes to a speedup of 75.8 (parallel efficiency 30\%) at 6,144 cores, normalized to the 24-core base. For the large scale problem (ball of 173,197,768 cells), JSweep shows a speedup of 9.9 (parallel efficiency of 62\%) at 49,152 cores, normalized to the 3,072-core base. Given that our tested mesh is a ball constructed with tetrahedrons, the above scalability should be reasonably good.

\subsubsection{Weak Scalability}
\begin{figure}[h]
	\centering
  	\includegraphics[scale = 0.25]{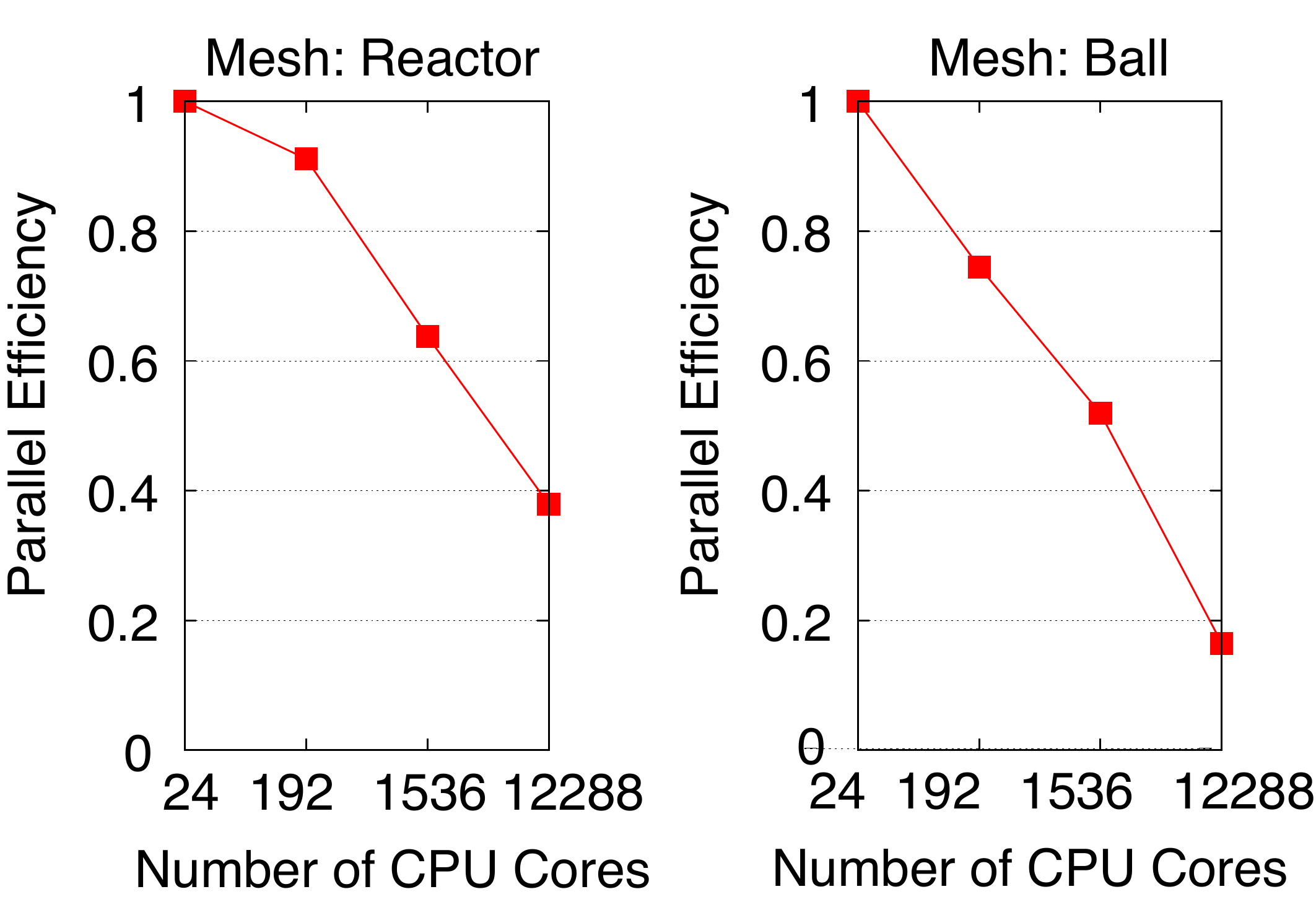}
\caption{Weak scalability of JSNT-U}
\label{fig:jsnt-u-weak}
\end{figure}

Fig.~\ref{fig:jsnt-u-weak} presents the results of weak scalability evaluated on the ball (originally 482,248 cells) and reactor core (originally 64,479 cells) meshes. In particular, mesh size is increased in a normal approximate refinement method. As shown, the weak scalability of JSweep is not good enough, although reasonable. For reactor, the parallel efficiency at 12,288 cores is about 40\%, while for ball it is lower than 20\%. One possible reason is that in JAxMIN, the original small mesh is first partitioned and distributed to processes, and then each process refines the assigned subdomain, leading to {\em thick} subdomains that dramatically increase length of critical path in the sweeping direction.

\subsection{Runtime Overhead Analysis}
\label{sub:profiling}
JSweep employs a runtime-based approach, thus the overhead is essential for performance. To investigate the overhead, we carry out a detailed profiling of JSNT-S on small scale Kobayashi benchmark. In particular, the problem has a $200 \times 200 \times 200$ mesh. All optimizations are enabled and all hyper parameters are the same with that in Sec.\ref{sub:evaluation-structured-mesh}. We present one sweep iteration using coarsened graph.

\begin{figure}[h]
	\centering
	\subfloat[Overall]{
  	\includegraphics[scale = 0.2]{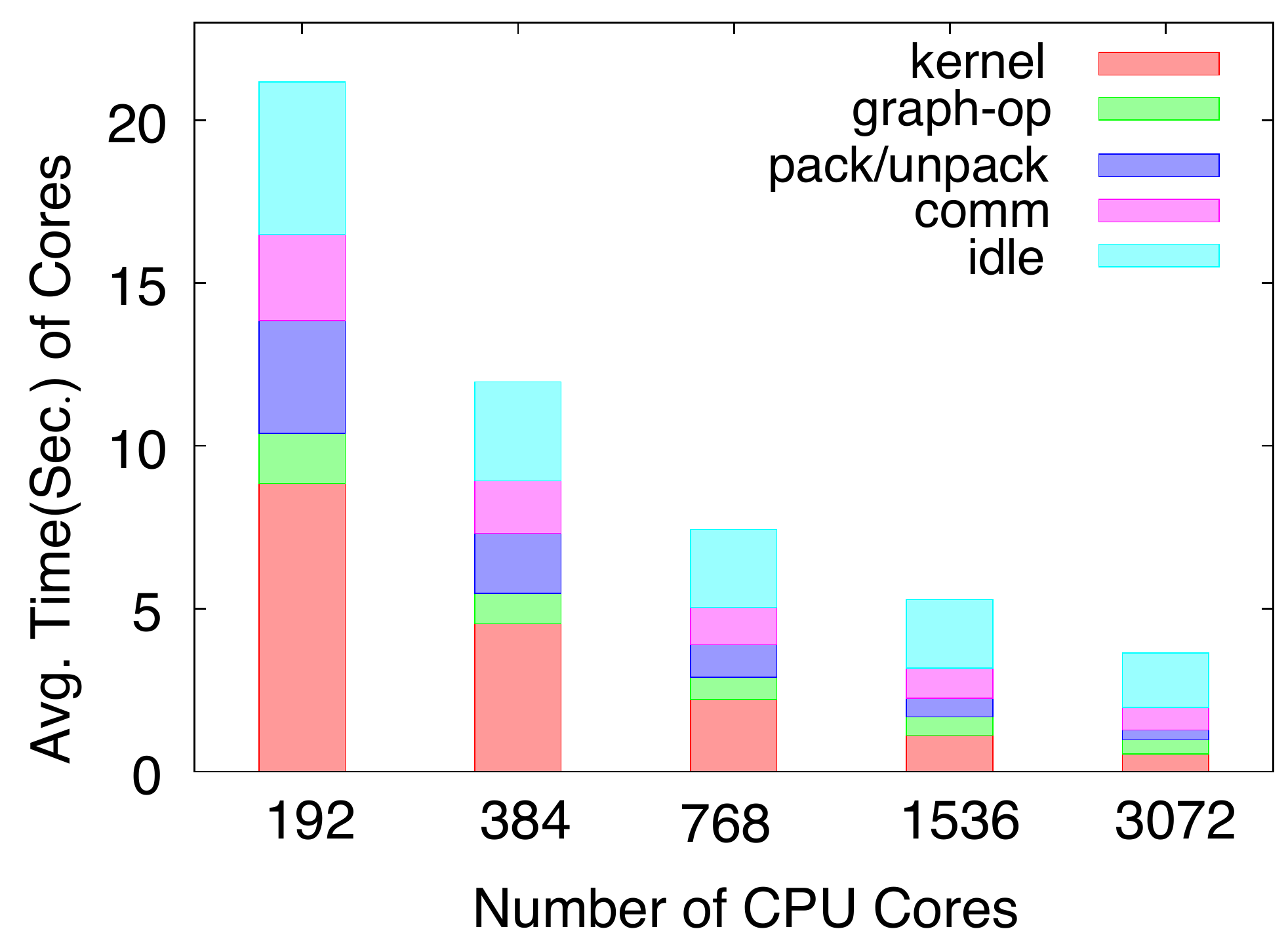}
  	\label{subfig:jsnt-s-time-break-all}
	}
	\subfloat[Workers]{
  	\includegraphics[scale = 0.2]{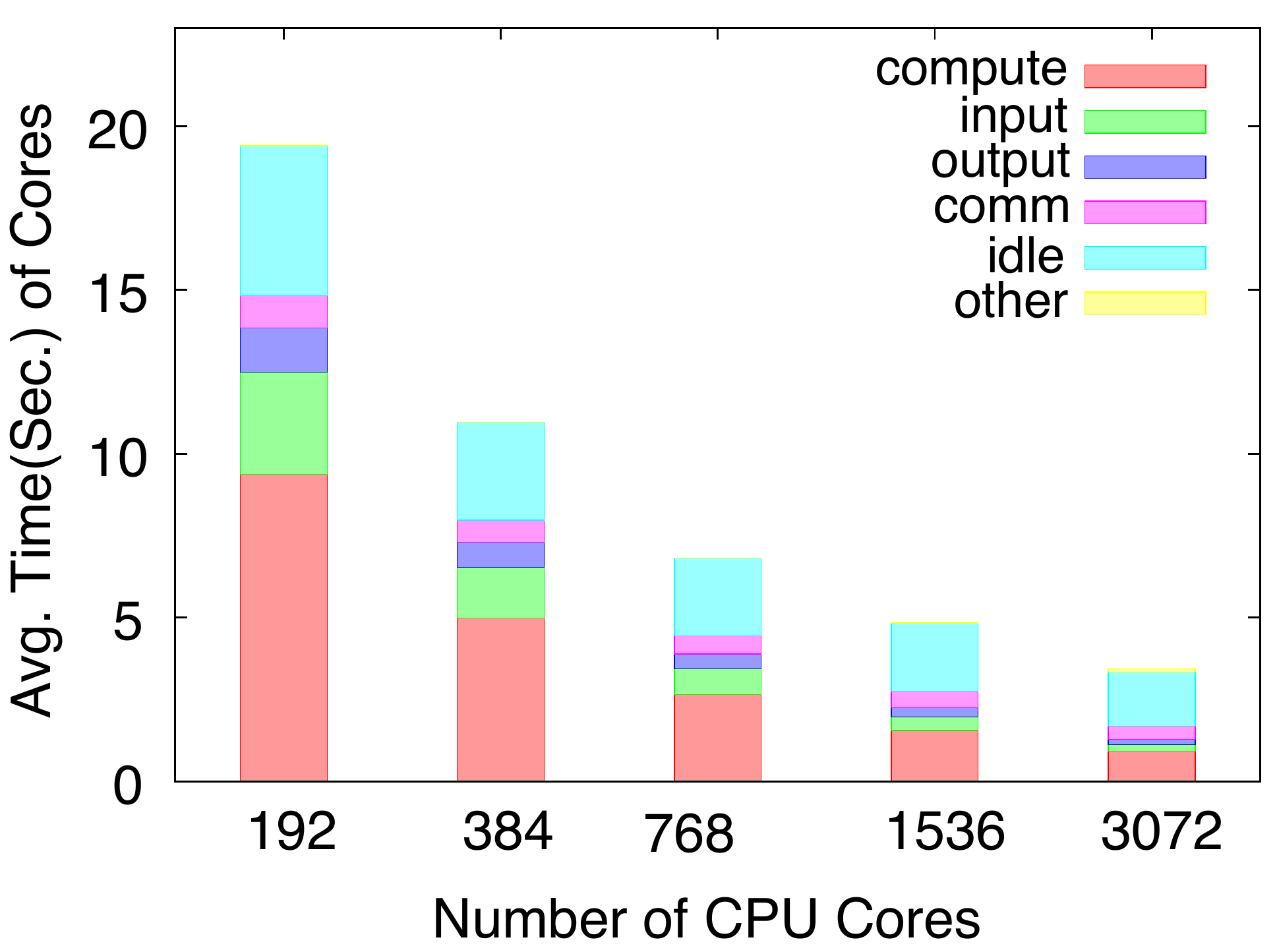}
  	\label{subfig:jsnt-s-time-break-worker}
	}
\caption{Runtime breakdown of JSNT-S}
\label{fig:jsnt-s-break-down}
\end{figure}

Fig.\ref{fig:jsnt-s-break-down} shows the time breakup in a strong-scaling fashion. The overhead introduced by JSweep (i.e., the graph-op and pack/unpack) is moderately low (approx. 23\%), and the major performance loss comes from idling of CPU cores (22\%-46\%). Communication takes 13\%-19\% the total time. With more deep optimization and advanced priority strategy, we expect to lower both the overhead and the idle time.
\vspace{-0.2cm}

\subsection{Performance Comparison with other systems}
\label{sub:comparison_others}

We first compare JSweep's performance with previous JAxMIN, which already implement efficient algorithmic optimizations ~\cite{mo2014} and achieve good performance. Nevertheless, we show that with innovations on data-driven abstraction and runtime system design, JSweep outperforms them on both structured and unstructured meshes.

\begin{figure}[h]
	\centering
	\subfloat[JSweep vs JASMIN (koba400)]{
  	\includegraphics[scale = 0.2]{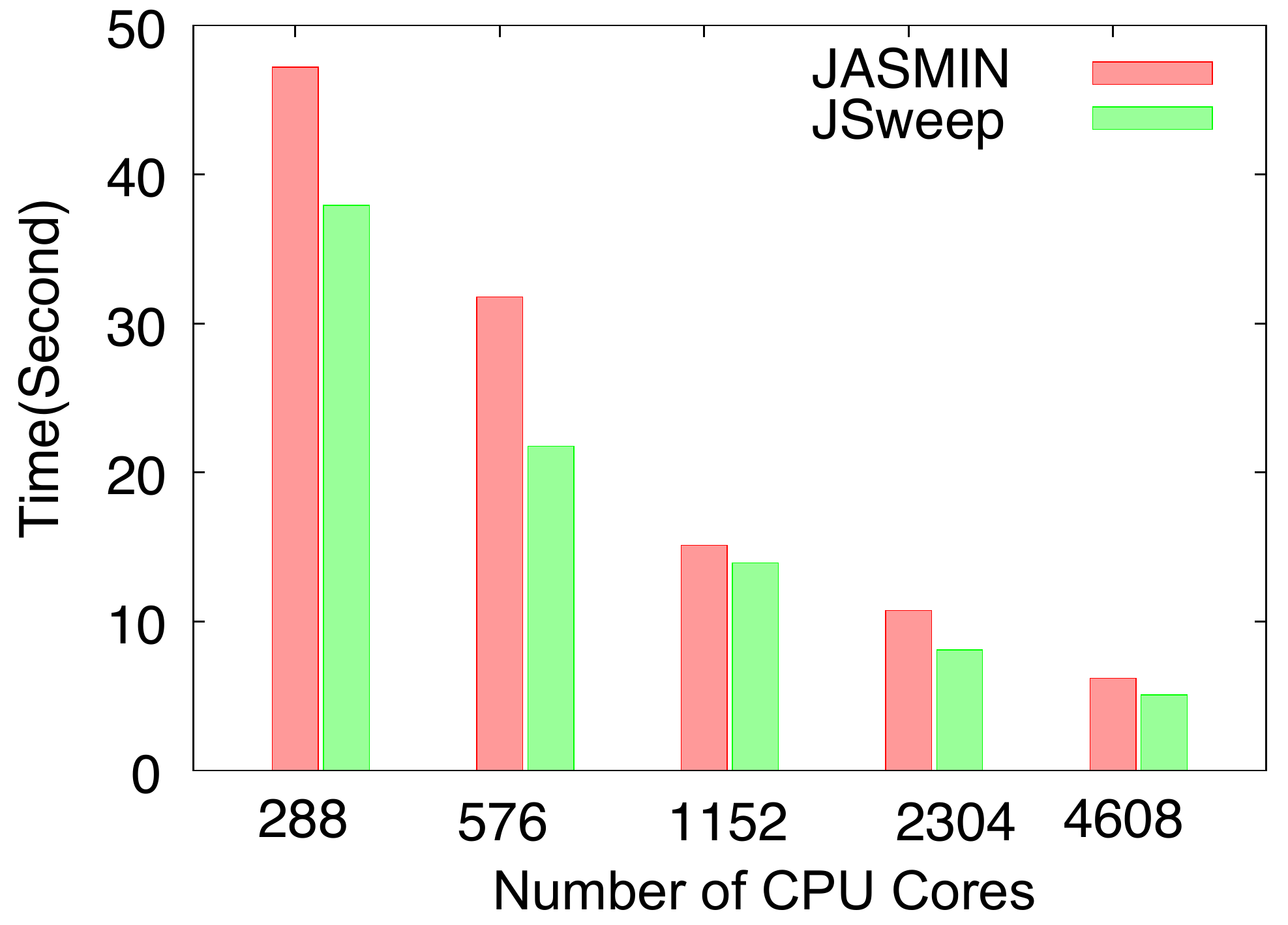}
  	\label{subfig:jsnt-s-jsweep-vs-jasmin}
	}
	\subfloat[JSweep vs JAUMIN (Ball)]{
  	\includegraphics[scale = 0.2]{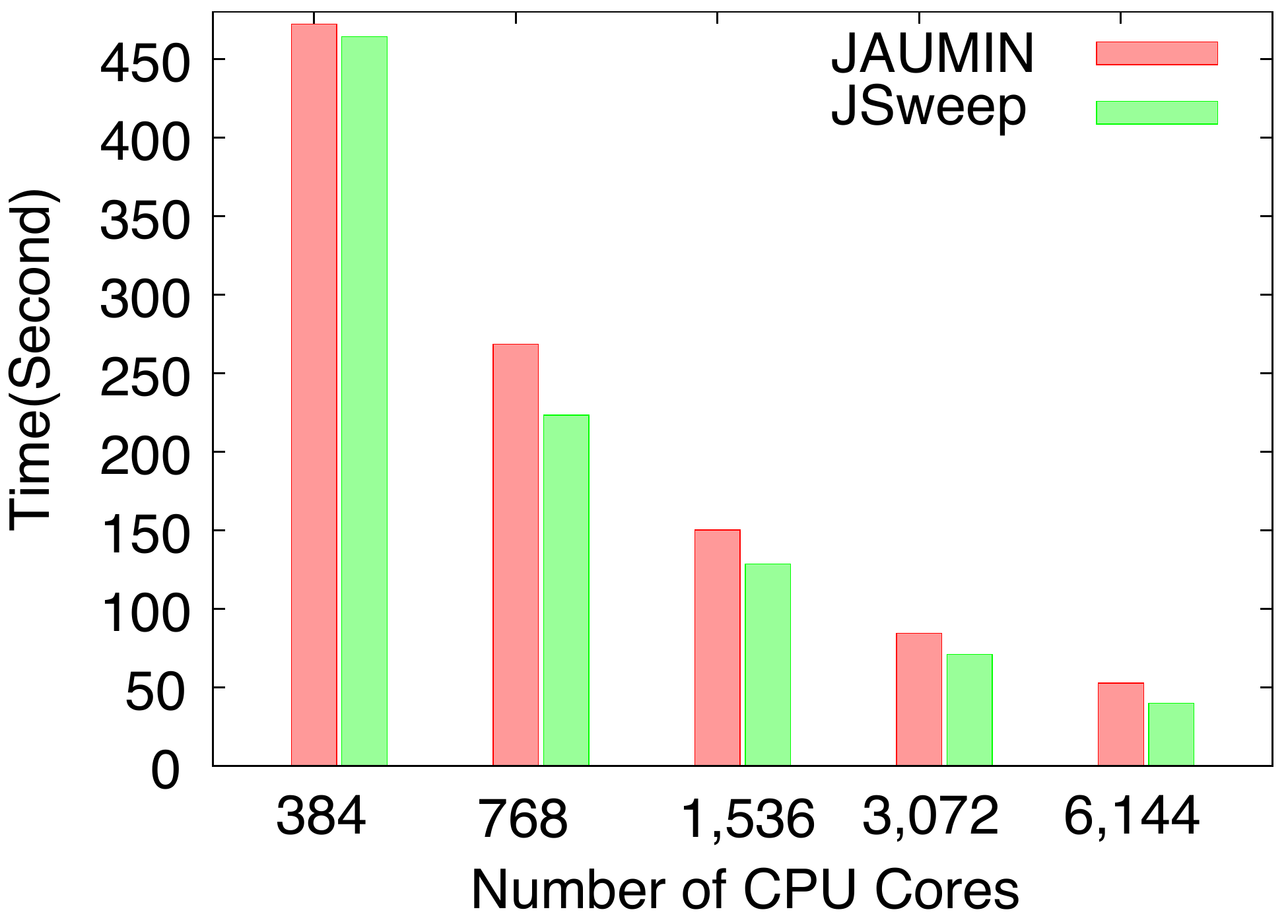}
  	\label{subfig:jsnt-u-jsweep-vs-jaumin}
	}
\caption{Performance comparison of JSweep vs JAxMIN}
\label{fig:jsnt-jsweep-vs-jaumin}
\end{figure}

Fig.\ref{subfig:jsnt-s-jsweep-vs-jasmin} presents results of JSweep and JASMIN-based SnSweep program (a data-driven implementation of Sweep3D\cite{sweep3d}). We choose SnSweep because it has been optimized manually with all techniques introduced in Sec.\ref{sec:sweep}, including a coarsened graph variant which caches the vertex clusters and their communication relationships by MPI tags. As shown, JSweep's runtime is constantly less than JASMIN.

Fig.\ref{subfig:jsnt-u-jsweep-vs-jaumin} presents runtime comparison of JSweep and JAUMIN-based JSNT-U. Again, JSweep shows constant runtime reduction, and with increasing number of cores the comparative advantage becomes slightly bigger.

Besides, Table-\ref{tab:perf-comparison} compares parallel efficiency of JSweep with other work in literatures. We can see that for Kobayashi problem, JSweep demonstrates comparable scalability with Denovo's KBA-based implementation. For unstructured ball (sphere) mesh of tetrahedrons, JSweep scales worse than the manually implemented data-driven algorithm PSD-b\cite{sn-colomer}. However, note that JSweep is a solution of general framework. Due to the lack of common public problems and availability of the systems, it is difficult to compare frameworks directly. 

\begin{table}[htpb]
\caption{Performance comparison with literatures}
\label{tab:perf-comparison}
\centering
{\small
\begin{tabular}{|l|c|r|l|}
\hline
Application & Problem & Par. Eff. & \#cores (max. vs base)\\
\hline
Denovo\cite{denovo-old} & Kobayashi-400 & 77.8\% & 3,600 vs 144\\
JSweep & Kobayashi-400 & 89.6\% & 6,144 vs 384\\
\hline
PSD-b\cite{sn-colomer} & sphere, 151,265, $S_{4}$ & ~88\% & 1,024 vs 128\\
JSweep & sphere, 482,248, $S_{4}$ & 66\% & 1,536 vs 192\\
\hline
\end{tabular}
}
\end{table}

\section{Related Work}
\label{sec:relatedwork}
The idea of patch-centric abstraction is partly inspired by the {\em vertex-centric} models~\cite{pregel} in graph-parallel frameworks~\cite{pregel}\cite{powergraph}. In a vertex-centric model, user defines a vertex-program for a single vertex and the framework lifts the vertex computation to the whole graph, conceptually in parallel. However, unlike vertex in graph, patch is not the basic element of mesh, which means patch-centric data-driven abstraction has fundamental difficulties including partial computation, priority inversion and multi-tasks on a single patch. In this paper we comprehensively addresses these issues and formalize a general patch-centric data-driven approach. In fact, the patch-centric abstraction can be seen as a straightforward extension to existing patch-based frameworks (see a survey in~\cite{frameworks-survey}), such as SAMRAI~\cite{samrai}, (part-based) PUMI~\cite{pumi} and especially JAxMIN\cite{jasmin}\cite{jaumin}, in which the mesh is decomposed into and managed by patches. 

Task-based programming models, such as PaRSEC~\cite{parsec} and more general Charm++~\cite{charmpp}, are also suitable to implement location-based data-driven computations. For example, a recent work~\cite{parsec-sweep} has implemented a PaRSEC-based $S_n$ sweep solver on 3d cartesian meshes, demonstrating high efficiency with 34\% of peak performance at 384 cores. Compared to JSweep, however, these task-based frameworks are not well-abstracted for mesh-specific parallel computation and thus require users to remap many conceptions.

\section{Conclusions and Future Work}
\label{sec:conclusion}
We have presented JSweep, a generic patch-centric data-driven framework integrated in the JAxMIN infrastructure. In particular, we propose the patch-centric data-driven abstraction whose essential idea is extending the concept {\em patch} as a logical processing element that is fully reentrant. Also, our abstraction supports multiple tasks on a single patch and arbitrary patch priority strategies. Further, targeting contemporary HPC systems of multicore cluster architecture, we implemented a high performance runtime system to map the patch-centric data-driven computation to underlying system resources. Based the above approach, we implemented a new parallel sweeps algorithm as a component in JAxMIN, featuring patch-angle parallelism, vertex clustering and hierarchical priority strategy. Evaluation with two real Sn software packages demonstrates that JSweep can scale to at least 49,152 cores for unstructured meshes and 76,800 cores for structured meshes with reasonable parallel efficiency.

What distinguishes JSweep from most counterparts is that we consider {\em sweep} computations in the context of general mesh-based application frameworks which would be critical in future coupling of multiple multi-physics simulations. Besides $S_n$ transport sweeps, our abstraction also supports other data-driven algorithms well, e.g., particle trace which we have implemented as another component in JAxMIN. Given the increasing importance of data-driven computation and demands on coupling multi-physics simulations, we believe efficient support to both BSP and asynchronous data-driven models are necessary to construct high performance applications.


\begin{thebibliography}{1}
\bibitem{jasmin}
Z. Mo, A. Zhang, X. Cao, Q. Liu, X. Xu, H. An, W. Pei, S. Zhu, JASMIN: a parallel software infrastructure for scientific computing, Frontiers Computer Science of China, 2010, 4(4):480--488. 

\bibitem{jaumin}
Q. Liu, W. Zhao, J. Cheng, Z. Mo, A. Zhang and J. Liu, A programming framework for large scale numerical simulations based on unstructured meshes, in {Proc. HPSC}, NewYork, Apr. 6--8, 2016.

\bibitem{top500}
http://top500.org/2015-nov, Nov., 2015.

\bibitem{boltzmann}
R. L. Bowers, J. R. Wilson, Numerical modeling in applied physics and astrophysics, Jones and Bartlett publishers, 1991.





\bibitem{dd-app-1-5}
J. Bey, G. Wittum, On the robust and efficient solution of convection diffusion problems on unstructured grids in two and three space dimensions, Applied Numerical Mathematics, 1997, 23(1):177--192.


\bibitem{dd-app-2-2}
F. Wang, J. Xu, A cross-wind strip block iterative method for convection-dominated problems, {\it SIAM Journal of Computing}, 1999, 21:646--665.


\bibitem{downar}
T. Downar, A. Siegel, C. Unal. Science Based Nuclear Energy Systems Enabled by Advanced Modeling and Simulation at the Extreme Scale. {\it White Paper on Integrated Performance and Safety Codes}, 2009.

\bibitem{t3d}
R. Baker, R. Alcouffe. Parallel 3-D Sn Performance for MPI on Cray-T3D. In {\it Proc. Joint International Conference on Mathematics Methods and Supercomputing for Nuclear Applications}, New York, Oct., 1997.

\bibitem{cm200}
R. Baker, K. Koch. An Sn algorithm for the massively parallel CM-200 computer. {\it Nuclear Science and Engineering}, 1998, 28: 312--320.

\bibitem{ardra}
Lawrence Livermore National Laboratory, Ardra: Scalable parallel code system to perform neutron and radiation transport calculations, http://www.llnl.gov/casc/ardra.


\bibitem{kba-improvement-2}
W. Hawkins, et al., Efficient Massively Parallel Transport Sweeps, {\it Trans. Am. Nucl. Soc.}, 107, 477, 2012.

\bibitem{sweep3d}
Los Alamos National Laboratory. The ASCI Sweep3d Benchmark. {\it http://www. ccs3.lanl.gov/pal/software/sweep3d.}

\bibitem{bsp}
G. Valiant. A bridging model for parallel computation. {\it Communications of the ACM}, 1990, 33(8):108--111.

\bibitem{consensus}
J. Misra. Detecting termination of distributed computations using markers. In {\em Proc. PODC}, pages 290--294, 1983.

\bibitem{plimpton-2000}
S. Plimpton, B. Hendrickson, S. Burns, W. McLendon. Parallel algorithms for radiation transport on unstructured grids. In {\it Proc. SC}, Dallas, Nov., 2000.


\bibitem{mo}
Z. Mo, A. Zhang and X. Cao. Towards a parallel framework of grid-based numerical algorithms on DAGs. In {\it Proc. IPDPS}, Greece, 2006.

\bibitem{dataflow}
Hewitt C, Bishop P, Steiger R. A Universal Modular Actor Formalism for Artificial Intelligence. In {\it Proc. IJCAI}, San Francisco, 1973.


\bibitem{metis}
G. Karypis, V. Kumar. Multi-level graph partitioning schemes. In {\it Proc. of ICPP}, Urbana-Champain, Aug., 1995, pp.113--122.

\bibitem{chaco}
B. Hendrickson, R. Leland. A multilevel algorithm for partitioning graph. In {\it Proc. of SC}, San Diego, 1995.


\bibitem{pautz}
S. Pautz. An Algorithm for Parallel Sn Sweeps on Unstructured Meshes. {\it Nuclear Science and Engineering}, 2002, 140(2): 111--136.


\bibitem{pmodel}
M. Mathis, D. Kerbyson. A General Performance Model of Structured and Unstructured Mesh Particle Transport Computations. {\it Journal of Supercomputing}, 2005, 34:181--199.




\bibitem{parsec-sweep}
S. Moustafa, M. Faverge, L. Plagne and P. Ramet. 3D Cartesian Transport Sweep for Massively Parallel Architectures with PaRSEC, In {\it Proc. IPDPS}, 2015.


\bibitem{frameworks-survey}
A. Dubey, et al., A survey of high level frameworks in block-structured adaptive mesh refinement packages. {\it J. Parallel Distrib. Comput.} (2014), http://dx.doi.org/10.1016/j.jpdc.2014.07.001.

\bibitem{tort}
W. A. Rhoades and D. Simpson, The TORT Three-Dimensional Discrete Ordinates Neutron/Photon Transport Code, ORNL/TM-13221, Oct., 1997.

\bibitem{jsnt-s}
T.P. Cheng and L. Deng, JSNT-S manual, IAPCM, 2015.


\bibitem{3dsn}
J.X. Wei,  JSNT-U (3DSn) manual, IAPCM, 2010.

\bibitem{sn-colomer}
G. Colomer, R. Borrell, F. Trias, and I. Rodrieguez, Parallel algorithms for Sn transport sweeps on unstructured meshes. {\it Journal of Computational Physics}, vol. 232, no. 1, pp. 118--135, 2013.

\bibitem{samrai}
SAMRAI. https://computation.llnl.gov/casc/SAMRAI, May 31, 2010

\bibitem{pumi}
D. A. Ibanez, E. S. Seol, C. W. Smith and Mark S. Shephard, PUMI: Parallel Unstructured Mesh Infrastructure. {\it ACM Transactions on Mathematical Software}, 2015.


\bibitem{denovo}
G. Davidson, T. Evans, J. Jarrell, S. Hamilton and T. Pandyam, Massively Parallel Three-dimensional Transport Solutions for the k-eigenvalue Problem, {\it Nuclear Science and Engineering}, 2014, 177:111-125.

\bibitem{denovo-old}
T. Evans, A. Stafford, R. Slaybaugh and K. Clarno, Denovo: a new three-dimensional parallel discrete ordinates code in scale. {\it Nuclear Technology}, 2010, 171(8), 171-200.


\bibitem{mo2014} 
Z. Mo, A. Zhang, and Z. Yang, A new parallel algorithm for vertex priorities of data flow acyclic digraphs, {\it The Journal of Supercomputing}, vol. 68, no. 1, pp. 49-€"64, 2014.

\bibitem{pautz-2015}
S. Pautz and T. Bailey, Parallel Deterministic Transport Sweeps of Structured and Unstructured Meshes with Overloaded Mesh Decompositions, {\it Proc. Joint International Conference on Mathematics and Computation (M\&C), Supercomputing for Nuclear Applications (SNA) and the Monte Carlo (MC) Methods}, Nashville, TN, April 19-23, 2015.



\bibitem{lared-all}
P. Song, C.L. Zhai, S.G. Lee, et.al, LARED-I: The Integrated Code for Laser-driven Inertial Confinement Fusion. {\it High Power Laser and Particle Beam} (in Chinese), 2015, 27(3):54-60.

\bibitem{multiphysics-reactor}
D. R. Gaston, C. J. Permann, J. W. Peterson, A. E. Slaughter, D. Andrsie, Y. Wang, M. P. Short, D. M. Perez, M. R. Tonks, J. Ortensi, Ling Zou and R. C. Martineau, Physics-based multi-scale coupling for full core nuclear reactor simulation, {\it Annals of Nuclear Energy}, 2015, 84:45--54.


\bibitem{parsec}
G. Bosilca, A. Bouteiller, A. Danalis, T. Herault, P. Lemarinier and J. Dongarra, DAGuE: A generic distributed DAG engine for High Performance Computing, {\it Parallel Computing}, vol.38, no.1-2, 2012.

\bibitem{charmpp}
L.V. Kale, E. Bohm, C. L. Mendes, T. Wilmarth and G. Zheng, Programming peta-scale applications with Charm++ and AMPI, {\it Peta-scale Computation: Algorithms Appl.} 1(2007): 421--441.

\bibitem{pregel}
G. Malewicz, M. Austern, A. Bik, J. Dehnert, I. Horn, N. Leiser and G. Czajkowski. Pregel: a system for large-scale graph processing. In {\it SIGMOD}, 2010.

\bibitem{powergraph}
J. Gonzalez, Y. Low, H. Gu, D. Bickson, C. Guestrin. PowerGraph: distributed graph-parallel computation on natural graphs. In {\it OSDI}, 2012.

\bibitem{graphine}
J. Yan, G. Tan, Z. Mo, N. Sun, Graphine: programming graph-parallel computation of large natural graphs for multicore clusters, In {\it IEEE Transaction on Parallel and Distributed Systems}, 27(6):1647-1659, 2016.

\end{thebibliography}
\end{document}